\documentclass[11 pt,oneside,onecolumn,a4paper]{article}
\usepackage{amsmath}
\usepackage{graphicx}
\DeclareGraphicsExtensions{.eps,.ps,.pdf}
\usepackage{bbm}
\usepackage{bm}
\usepackage{epsfig}
\usepackage[T1]{fontenc}
\usepackage{esint}
\usepackage{amssymb}

\usepackage{lipsum}
\usepackage{savesym}

mm \leftmargin 5pt \rightmargin 5pt \hoffset= -15mm

\title{Short-time coherence of a qubit and measurement apparatus}
\author{Filippo Giraldi}

\date{\small{School of Chemistry and Physics, University of KwaZulu-Natal\\
Westville Campus, Durban 4000, South Africa
}}

\begin{document}

\maketitle

\begin{abstract}
The effects of the measurement apparatus on quantum coherence are studied by considering a purely dephasing model of a qubit. The initial state is prepared from a thermal state of the whole system by performing a nonselective measurement on the qubit. The magnitude of the initial postmeasurement coherence is bounded by the value $1/2$, which is realized with special measurement schemes and in the low-temperature limit. The magnitude of coherence identically vanishes, increases or decreases with approximately constant velocity over a determined short time scale, according to the choice of the preparation measurement. The maximization of the short-time increasing or decreasing velocity is favored by the choice of further special measurement schemes and the high-temperature limit. The measurement apparatus allows to manipulate quantum coherence of the qubit over short times via nonselective preparation measurements.  
\end{abstract}

PACS: 03.65.Yz, 03.65.Ta



\section{Introduction}\label{1}

The measurement process which is performed on an open quantum system, for the preparation of the initial state, can influence the reduced dynamics of both the open system and the external environment \cite{BP,measure1,measure2,measure3,measure4,measure5,measure6,measure7,M0,M1,M2,M3,G_IJQI2020}. This behavior is caused by the correlations which initially exist between the open quantum system and the external environment. The reduced dynamics of a qubit and the bosonic environment is analyzed in Refs. \cite{M0,M1,M2,M3} by considering a purely dephasing model which is exactly solvable \cite{LeggettRMP1987,RH1}. The initial state is prepared from the thermal state of the whole system by performing a selective or nonselective measurement on the qubit. Selective measurements prepare the qubit in a pure initial state, while nonselective measurements prepare the whole system in a mixed initial state \cite{M0,M1,M2,M3}. Usually, the preparation with selective measurements tends to enhance the decoherence process of the qubit. Instead, quantum coherence can increase significantly over intermediate times in case the initial condition is prepared with appropriate nonselective measurement schemes (NMSs) \cite{M0,M1,M2,M3}. Quantum coherence can increase or decrease over short times \cite{G_IJQI2020}. The initial increasing or decreasing behavior is determined by the NMS and depends on the spectral density (SD) of the system. This function is determined by the couplings among the qubit and the bosonic modes of the external environment.

Decoherence-recoherence events consist in a decreasing, vanishing and, then, an increasing behavior which appear in the evolution of coherence. By considering the purely dephasing model \cite{LeggettRMP1987,RH1}, decoherence-recoherence events appear in the reduced dynamics of the qubit uniquely if special nonselective preparation measurements (NPMs) are adopted and the SD exhibits special properties \cite{GOSID2017}. The number of decoherence-recoherence events can be finite or infinite and can be estimated for canonical ohmic-like environments. This number is finite for super-ohmic and ohmic SDs, while it is infinite for sub-ohmic SDs. Furthermore, the properties of NPMs and the SD determine the decoherence-recoherence times and allow to relate the variation of the environmental energy to the flow of quantum information \cite{GOSID2017}.

As a continuation of the above-described scenario, here, we rely on the analysis which is performed in Refs. \cite{M0,M1,M2,G_IJQI2020} and we study how the measurement apparatus affects quantum coherence of the qubit by considering every configuration of the NMS. Particularly, we intend to study the effects of the NPM on the initial postmeasurement coherence and the increase or decrease of quantum coherence in the short-time evolution. This analysis aims to provide conditions on the setting of the measurement apparatus which allow to maximally enhance of coherence of a qubit, or 
avoid maximally reduction, via the properties of the NMSs. The description and the implementation of the measuring apparatus are beyond the purposes of the present paper. 

 The paper is organized as follows. Section \ref{2} is devoted to the description of the model and the NPMs. The short-time evolution of quantum coherence of the qubit is described in Section \ref{3}. The maximal magnitude of the initial postmeasurement coherence is determined in Section \ref{4} along with the NMSs which realize this condition. Section \ref{5} is devoted to the maximal short-time increase and decrease of quantum coherence and to the NPMs which induce these conditions. The transition from full decoherence to the maximally decreasing or to maximally increasing regime is analyzed in the same Section. Summary of the results and conclusions are provided in Section \ref{6}. Details of the calculations are given in the Appendix.

\section{The model}\label{2}

The model under study consists in a qubit (two-level system) which interacts with a bosonic environment \cite{BP,M0,M1,M2,M3,LeggettRMP1987,RH1}. The Hamiltonian $H$ of the whole system is given by the sum of three terms, $H=H_S+H_{SE}+H_E$, which are defined by the forms below,
\begin{eqnarray}
&&\hspace{-1em}H_S=\frac{\omega_0}{2} \sigma_z,\hspace{1em}
H_E=\sum_k \omega_k b^{\dagger}_k b_k, 
\hspace{1em}H_{SE}=\sigma_z\otimes\sum_k\left(g_k  b^{\dagger}_k+g^{\ast}_k b_k\right). \label{HSE}
\end{eqnarray}
In the present system of units the Planck constant $\hbar$ and the Boltzmann constant $k_B$ are equal to unity, $\hbar=k_B=1$. The term $H_S$ is the microscopic Hamiltonian of the qubit and acts on the corresponding Hilbert space $\mathcal{H}_S$. The operator $\sigma_z$ is the $z$-component of the Pauli spin operator \cite{BP,LeggettRMP1987}. The ket $|0\rangle$ is the ground state of the Hamiltonian $H_S$ of the qubit, while the ket $|1\rangle$ is the excited state, and $\omega_0$ represents the transition frequency, or, equivalently, the magnitude of the energy difference between the two eigenstates. The term $H_E$ is the Hamiltonian of the bosonic environment and acts on the corresponding Hilbert space $\mathcal{H}_E$. The rising operator $b^{\dagger}_k$ and the lowering operator $b_k$ act on the Hilbert space of the bosonic mode with frequency $\omega_k$. The term $H_{SE}$ of the Hamiltonian mimics the linear interaction between the qubit and the external environment. The constant $g_k$ quantifies the coupling between the open system and the bosonic mode with frequency $\omega_k$. This notation holds for every value of the index $k$.

\subsection{Nonselective preparation measurements}\label{21}

The preparation of the initial state of the qubit via a nonselective measurement is described below, for the sake of clarity, by following Ref. \cite{M0}. Before the measurement process, the whole system is initially in a thermal state at temperature $T$ and is described by the following density matrix, $\exp\left(-H/T\right)/Z_0$. The normalization constant $Z_0$ is obtained by performing the following trace operation over the degrees of freedom of the whole system, $Z_0=\operatorname{Tr} \left\{\exp \left(-H/T\right)\right\}$. In this condition, the qubit is correlated with the bosonic environment. At this stage, an observable with discrete, nondegenerate spectrum is measured \cite{M0}. Due to the measurement process, the whole system is induced in the mixed state $\rho_T(0)$ which is given by the expression below \cite{measure1,measure2,measure3,measure4,measure5,measure6},
\begin{eqnarray}
\rho_T(0)=\frac{1}{Z_0}\sum_j O_j \exp\left(-\frac{H}{T}\right)O^{\dagger}_j.
\label{rhoTOj}
\end{eqnarray}
For every value of the index $j$, the operator $O_j$ acts on the Hilbert space of the open system and mimics the $j$th outcome of the measurement, $O_j= \left|\varphi_j\rangle\langle\psi_j\right|$. In fact, the state $|\varphi_j\rangle$ represents the transformed of the state $|\psi_j\rangle$ which belongs to an orthonormal base of the open system. The probability distribution $p_j$ that the measure of the observable provides the $j$th outcome is given by the following form, $p_j= \operatorname{Tr}\left\{O^{\dagger}_j O_j\exp\left(-H/T\right)/Z_0\right\}$. The present description refers to the condition where the effects, $F_j=O^{\dagger}_j O_j$, are projectors, $F_j=\left|\psi_j\rangle\langle\psi_j\right|$ and $F_j^2=F_j$ for every value of the index $j$. This scenario belongs to the framework of the Neumann-L\"uder projection measurement \cite{measure1,measure2,BP,M0}.

In the present case, the nonselective preparation measurement is performed over a qubit. The operators $O_1$ and $O_2$, appearing in Eq. (\ref{rhoTOj}), are defined in terms of the eigenstates of the Pauli spin operators as follows. Let $\sigma_{\hat{n}}$ be the component of the Pauli spin operator in the direction which is given by the three-dimension unit vector $\hat{n}$. The general eigenstate $|\hat{n}\rangle$ of the spin operator $\sigma_{\hat{n}}$ with the eigenvalue $1$, i.e., $\sigma_{\hat{n}}|\hat{n}\rangle=|\hat{n}\rangle$, is given by the form below,
\begin{eqnarray}
|\hat{n}\rangle=\exp\left(-\imath \frac{\zeta}{2}\right)\cos\frac{\vartheta}{2} |1\rangle+\exp\left(\imath \frac{\zeta}{2}\right)\sin\frac{\vartheta}{2} |0\rangle,
\label{ketsigman}
\end{eqnarray}
where $\vartheta$ and $\zeta$ are the polar angle and the azimuthal angle of the unit vector $\hat{n}$, respectively. Beside the orthogonality property, the kets $|\hat{n}\rangle$ and $|-\hat{n}\rangle$ fulfill the following completeness condition: $|\hat{n}\rangle\langle \hat{n}|
+|-\hat{n}\rangle\langle -\hat{n}|=I$, where $I$ is the identity operator. The operators $O_1$ and $O_2$ read as follows: $O_1=|\hat{n}_1\rangle\langle \hat{n}_0|$ and $O_2=|\hat{n}_2\rangle\langle -\hat{n}_0|$. The three states $|\hat{n}_0\rangle$, $|\hat{n}_1\rangle$ and $|\hat{n}_2\rangle$ are eigenstates, with eigenvalue $1$, of the components of the Pauli spin operator in the directions which are given by the three-dimension unit vector $\hat{n}_0$, $\hat{n}_1$ and $\hat{n}_2$, respectively. This means that the following equality, $\sigma_{\hat{n}_j}|\hat{n}_j\rangle=|\hat{n}_j\rangle$, holds for every $j=0,1,2$. The postmeasurement states $|\hat{n}_1\rangle$ and $|\hat{n}_2\rangle$ represent the action of the measuring apparatus on the basis states $|\hat{n}_0\rangle$ and $|-\hat{n}_0\rangle$. For example, the condition $\hat{n}_1=\hat{n}_0$ and $\hat{n}_2=-\hat{n}_0$ describes nonselective measurements which do not affect the basis states. In this case the operators $O_1$ and $O_2$ read $O_1=|\hat{n}_0\rangle \langle \hat{n}_0|$ and $O_2=|-\hat{n}_0\rangle \langle -\hat{n}_0|$. Refer to \cite{M0} for details.

\subsection{Quantum coherence} \label{22}

At every time instant $t$ the qubit is described by the reduced density matrix $\rho(t)$ which is obtained by performing the trace operation on the density matrix of the whole system, $\rho_T(t)$, over the degrees of freedom of the external environment,
\begin{eqnarray}
\rho(t)=\operatorname{Tr}_{\rm E}\left\{\exp\left(-\imath H t\right) \rho_T(0)\exp\left(\imath H t\right)\right\}. \label{rhoSt}
\end{eqnarray}
The off-diagonal elements of the reduced density matrix $\rho(t)$ represent the coherence which exists between the state $|0\rangle$ and $|1\rangle$ of the qubit, $\rho_{0,1}(t)=\langle 0|\rho(t)|1\rangle=\langle \sigma_+(t)\rangle=\langle \sigma_-(t)\rangle^{\ast}=\rho_{1,0}^{\ast}(t)$. The Pauli matrices $\sigma_x$ and $\sigma_y$ define the rising and lowering operators $\sigma_+$ and $\sigma_-$, respectively, as follows: $\sigma_{\pm}=\left(\sigma_x\pm \imath \sigma_y\right)/2$.

Let the whole system be in a thermal state and let the initial condition be prepared by performing a nonselective measurement on the qubit. Following Ref. \cite{M0}, the time evolution of the coherence term, $\rho_{0,1}(t)$, is given by the expression below,
\begin{eqnarray}
&&\hspace{-4.0em}\rho_{0,1}(t)=\rho_{0,1}(0)
\exp\left\{\imath \left[\omega_0 t+\chi_{1,T}(t)\right]-\Xi_T(t)\right\}
\nonumber \\ && \times 
\sqrt{1+a_{2,T} \sin \left[2\upsilon_0(t)\right]+a_{3,T} \sin^2 \left[\upsilon_0(t)\right]}. \label{sigmapmtNNselect}
\end{eqnarray}
The constants $a_{2,T}$ and $a_{3,T}$ are defined in the Appendix. The value $\rho_{0,1}(0)$ of the initial postmeasurement coherence term is determined by the NPM \cite{M0},
\begin{eqnarray}
&&\hspace{-1.5em}\rho_{0,1}(0)=
\frac{\exp \left(\imath \zeta_1\right)}{4 \cosh\left[\omega_0/\left(2T\right)\right]}\Bigg\{\sin \vartheta_1\left[\exp\frac{\omega_0}{2T}\sin^2 \frac{\vartheta_0}{2}+\exp\left(-\frac{\omega_0}{2T}\right)\cos^2 \frac{\vartheta_0}{2}\right]
\nonumber \\ && \hspace{2.8em}+
\exp \left(-\imath \Delta_{\zeta}\right)\sin \vartheta_2
\left[\exp\frac{\omega_0}{2T}\cos^2 \frac{\vartheta_0}{2}+\exp\left(-\frac{\omega_0}{2T}\right)\sin^2 \frac{\vartheta_0}{2}\right]
\Bigg\}.
  \label{rho01t0NNselect}
\end{eqnarray}
The polar angle $\vartheta_j$ and the azimuthal angle $\zeta_j$ characterize the three-dimensional unit vector $\hat{n}_j$, for every $j=0,1,2$. The function $\Xi_T(t)$ represents the dephasing factor and depends on the temperature $T$ and the coupling between the system and the environment,
 \begin{eqnarray}
\Xi_T(t)=\int_{\omega_g}^{\omega_M}J_T\left(\omega\right)\frac{1-\cos\left(\omega t\right)}{\omega^2} \, d \omega. \label{XiTt}
\end{eqnarray}
The effective SD $J_T\left(\omega\right)$ is defined as follows: $J_T\left(\omega\right)=J\left(\omega\right) \coth \left[\omega/\left(2 T\right)\right]$, for every non-vanishing value of the temperature $T$. The function $J\left(\omega\right)$ is the SD of the system. This function is fundamental for the description of the reduced dynamics of the qubit and depends on the coupling constants as follows, 
\begin{eqnarray}
J\left(\omega\right)=\sum_k \left|g_k\right|^2 \delta \left(\omega-\omega_k\right). \label{SDdef}
\end{eqnarray}
The SD encodes the properties of the bosonic environment and the interaction between open system and external environment. For a continuous distribution of frequency modes the SD is defined in terms of the mode frequency $\omega$ and the density of modes $r \left(\omega\right)$ as follows \cite{BP,SDPlenio1,M0,M1,M2,M3}:  
\begin{eqnarray}
J\left(\omega\right)= 4 r \left(\omega\right) \left|g\left(\omega\right)\right|^2. \label{SDcontModes}
\end{eqnarray}
The effective SD $J_T\left(\omega\right)$ reproduces the ordinary SD $J\left(\omega\right)$ for vanishing temperature, $J_T\left(\omega\right) \to J\left(\omega\right)$ as $T \downarrow 0^+$, for every $\omega\geq\omega_g$. The frequency $\omega_g$, appearing in Eq. (\ref{XiTt}), represents the upper cutoff frequency in case the continuous distribution of frequency modes exhibits a low-frequency band gap. However, the present analysis includes the condition where no low-frequency band gap exists in the distribution of frequency modes by setting $\omega_g=0$. The maximum mode frequency $\omega_M$ can be either finite or infinite. The auxiliary function $\upsilon_0(t)$, appearing in Eq. (\ref{sigmapmtNNselect}), is defined by the following form:
\begin{eqnarray}
&&\upsilon_0(t)=\int_{\omega_g}^{\omega_M} J\left(\omega\right) \frac{\sin\left(\omega t \right)}{\omega^2} \, d\omega.
 \label{upsilon0tdef}
\end{eqnarray}
The time-dependent phase shift $\chi_{1,T}(t)$ is defined via the auxiliary function $\upsilon_0(t)$ by the expression below \cite{M0},
\begin{eqnarray}
&&\tan \left[\chi_{1,T}(t)\right]=\frac{ N_{1,T} \sin \left[\upsilon_0(t)\right]}{N_{0,T}\cos \left[\upsilon_0(t)\right]+N_{2} \sin \left[\upsilon_0(t)\right]}. \label{chi_1tdef} 
\end{eqnarray}
The constants $N_{0,T}$, $N_{1,T}$ and $N_{2}$ are defined in the Appendix.

If the initial state of the qubit is decoupled from the thermal state of the bosonic environment, at temperature $T$, the coherence term $\rho_{0,1}(t)$ of the reduced density matrix of the qubit evolves as below \cite{BP,RH1},
\begin{equation}
\rho_{0,1}(t)=\rho^{\ast}_{1,0}(t)=\rho_{0,1}(0)\exp \left[-\Xi_T(t)\right].  \label{rho01tTdecoupled}
\end{equation}
In Ref. \cite{M0}, among many other relevant results, it is shown how the initial correlations, which are created by the nonselective preparation measurements, can enhance quantum coherence. In fact, beside the time-dependent phase $\exp\left\{\imath \left[\omega_0 t+\chi_{1,T}(t)\right]\right\}$, the following factor: \\$\sqrt{1+a_{2,T} \sin \left[2\upsilon_0(t)\right]+a_{3,T} \sin^2 \left[\upsilon_0(t)\right]}$ is added to the canonical decoherence law (\ref{rho01tTdecoupled}) in case a NPM is performed. This term is larger than unity in case the parameter $N_{2,T}$ vanishes, $N_{2,T}=0$, and the ratio $N^2_{1,T}/N^2_{0,T}$ is larger than unity, $N^2_{1,T}>N^2_{0,T}$. Hence, in these conditions, the canonical value of quantum coherence is enhanced. For example, the NMSs which correspond to the conditions $\vartheta_1+\vartheta_2=\pi$, $\sin \Delta_{\zeta}=0$, $\cos\Delta_{\zeta}=-1$, or to the conditions $\vartheta_1=\vartheta_2$, $\sin \Delta_{\zeta}=0$, $\cos\Delta_{\zeta}=-1$, fulfill the above-mentioned constraints and enhance the canonical form (\ref{rho01tTdecoupled}) of quantum coherence. The angle $\Delta_{\zeta}$ is defined as the difference of the azimuth angles $\zeta_1$ and $\zeta_2$, i.e., $\Delta_{\zeta}=\zeta_1-\zeta_2$, and belongs to the interval $\left.\right]-2\pi,2\pi\left[\right.$, as $0\leq\zeta_j<2\pi$ for every $j=0,1,2$. Qualitatively, quantum coherence increases if the postmeasurement states are close to the states which are involved in the above-mentioned nonselective measurement schemes, $\cos\Delta_{\zeta}=-1$, and $\vartheta_1+\vartheta_2\simeq\pi$, or $\vartheta_1\simeq\vartheta_2$. Refer to \cite{M0} for details.


\section{Short-time evolution of coherence 
}\label{3}

In case the initial condition is prepared with a nonselective measurement, the evolution of the environmental energy and of the reduced density matrix of the qubit are determined by the SD, the temperature $T$ of the initial thermal state of the whole system and by the features of the NMS \cite{M3,M0,M1,M2}. The short- and long-time evolutions of quantum coherence are studied in Refs. \cite{M0,M1,M2}, for ohmic-like SDs, and in Ref. \cite{G_IJQI2020}, in case the distribution of frequency modes exhibits a bang gap over low frequencies. The magnitude of the coherence term tends to the following asymptotic value: 
\begin{eqnarray}
&&\hspace{-2em}\left|\rho_{0,1}\left(\infty\right)\right|=\left|\rho_{0,1}(0)\right|
\exp \left[- 
\int_{\omega_g}^{\omega_M}\frac{ J_T\left(\omega\right)}{\omega^2}\,
d \omega \right], \label{rhoinftygapT}
\end{eqnarray}
in case the initial condition is factorized or it is prepared with selective or nonselective measurements. In absence of the spectral gap, $\omega_g=0$, the initial coherence is totally lost in case the quantity $\int_0^{\omega_M} J_T\left(\omega\right)/\omega^2\,d \omega$ is infinite, while residual coherence persists over long times if this quantity is finite. The presence of a low-frequency band gap maintains a residual amount of coherence over long times as the quantity $\int_{\omega_g}^{\omega_M} J_T\left(\omega\right)/\omega^2\,d \omega$ is  positive and finite for canonical forms of the SD. Let the following scaling property, $J \left( \omega_g+ \omega_s\nu \right)=\omega_s\Omega\left(\nu \right)$, hold for every $\nu\geq 0$. The parameter $\omega_s$ is a typical scale frequency of the system, while $\Omega\left(\nu\right)$ is a dimensionless auxiliary function. The long time scale is determined by the scale frequency $\omega_s$. In fact, over long times, $t \gg 1/\omega_s$, the magnitude of the coherence term, $\left|\rho_{0,1}\left(t\right)\right|$, tends to the asymptotic value, $\left|\rho_{0,1}\left(\infty\right)\right|$, according to damped oscillations which are determined by the low-frequency band gap. The envelope of the damped oscillations is determined by the behavior of the SD near the low-frequency gap. By considering initial conditions which are factorized or are prepared with selective or nonselective measurements, the last preparation procedure provides the slowest among the long-time decaying envelopes of the oscillations \cite{G_IJQI2020}.

Differently from the long-time behavior, the short-time coherence increases, decreases or vanishes according to the features of the NMS. This property suggests that quantum coherence can be manipulated, over short times, via the NPM \cite{M3,M0,M1,M2,G_IJQI2020}. For the sake of clarity, the short-time evolution of the modulus of the coherence term of the reduced density matrix of the qubit is described below in terms of the properties of the NMS. Let the SD decay sufficiently fast at high frequencies in such a way that the following constraint holds, 
\begin{eqnarray}
\int_{\omega_g}^{\omega_M}\frac{\omega^2}{\omega^4_s} J_T\left(\omega\right) \,d \omega<\infty. \label{SDfastdecay1}
\end{eqnarray}
 Under this condition, the modulus of the coherence term evolves approximately algebraically over short times if the initial state of the qubit is prepared with a nonselective measurement \cite{G_IJQI2020},
\begin{eqnarray}
\left|\rho_{0,1}(t)\right|\sim \left|\rho_{0,1}(0)\right|+ \mathfrak{V}t +
 \mathfrak{W}t^2,
\label{Absrho01shortNNselect}
\end{eqnarray}
for $t \ll t_{\rm s}$. The coefficient $\mathfrak{V}$ of the linear term and the coefficient $\mathfrak{W}$ of the quadratic term are determined by the SD, the temperature $T$ and the NMS. The corresponding expressions are reported in the Appendix. The short times, $t \ll t_{\rm s}$, depend on the time scale $t_{\rm s}$ which is determined uniquely by the SD and by the temperature $T$,
\begin{eqnarray}
t_{\rm s} = \min \left\{ \frac{1}{\omega_s},\sqrt{ 6\frac{\eta_{-1,0}}{\eta_{1,0}}},2\sqrt{3 \frac{\eta_{0,T}}{\eta_{2,T}}}\right\}. \label{ts}
\end{eqnarray}
The parameters, $\eta_{-1,T}, \ldots, \eta_{2,T}$, are defined in the Appendix. 
According to Eq. (\ref{Absrho01shortNNselect}), a short time scale exists, $t \ll t_{\rm l}$, over which the modulus of the coherence term, $\left|\rho_{0,1}(t)\right|$, evolves approximately linearly,
\begin{eqnarray}
\left|\rho_{0,1}(t)\right|\sim \left|\rho_{0,1}(0)\right|+ \mathfrak{V}t.   \label{Absrho01shortNNselectLINEAR}
\end{eqnarray}
The time scale $t_{\rm l}$ is determined by the SD, the temperature $T$ and, additionally, by the NMS \cite{G_IJQI2020},
\begin{eqnarray}
t_{\rm l}= \min \left\{\frac{1}{\omega_s},\sqrt{ 6\frac{\eta_{-1,0}}{\eta_{1,0}}},2\sqrt{3 \frac{\eta_{0,T}}{\eta_{2,T}}},2\eta_{-1,0}\left|\frac{a_{2,T}}{\xi_{T}}\right| \right\}.\label{tl}
\end{eqnarray}
The parameter $\xi_{T}$ is defined in the Appendix. Notice that the following relations hold about the time scales: $t_{\rm l}\leq t_{\rm s}\leq 1/\omega_s$. The time scales $t_{\rm s}$ and $t_{\rm l}$ are properly defined in absence of the low-frequency gap, $\omega_g=0$. In fact, besides the required constraint (\ref{SDfastdecay1}), the parameter $\eta_{-1,0}$ is finite for $\omega_g=0$ \cite{SDPlenio1,ReedSimonBook}. 

In the present dephasing model, coherence of the qubit identically vanishes during the whole reduced evolution, $\left|\rho_{0,1}(t)\right|=0$ for every $t \geq 0$, if the initial postmeasurement coherence vanishes, $\left|\rho_{0,1}(0)\right|=0$. This condition is realized if every postmeasurement state coincides with any of the states $|1\rangle$ and $|0\rangle$. Coherence vanishes during the whole reduced evolution also in case the initial state of the qubit is prepared with special NMSs which are labeled, here, as $\mathfrak{S}_0$. The variety of polar angles which characterize the postmeasurement states of these special schemes is described by the parameter $q$. This parameter is defined as below,
\begin{eqnarray}
q\equiv \frac{\sin \vartheta_1}{\sin \vartheta_2},
\label{qdef}
\end{eqnarray}
in case the polar angle $\vartheta_2$ belongs to the interval $ \left.\right] 0,\pi\left[\right.$. The special NMSs $\mathfrak{S}_0$ are defined as follows:
\begin{eqnarray}
&&\hspace{-1em}q=Q\left(\frac{\omega_0}{T}, \vartheta_0\right),
\hspace{1em}\Delta_{\zeta}=\pm \pi, \hspace{1em}\forall \,\, \vartheta_0\in 
\left[0,\pi\right], \hspace{1em}\forall \,\,\zeta_0 \in \left[\right. 0,2\pi\left[\right..\label{FDNMS1}
\end{eqnarray}
The special value $Q\left(\omega_0/T, \vartheta_0\right)$ of the parameter $q$ is given by the form below,
\begin{eqnarray}
Q\left(\frac{\omega_0}{T}, \vartheta_0\right)=\frac{1+\exp\left(-\omega_0/T\right)+\left[1-\exp\left(-\omega_0/T\right)\right]\cos \vartheta_0}{1+\exp\left(-\omega_0/T\right)-\left[1-\exp\left(-\omega_0/T\right)\right]\cos \vartheta_0},
\label{QDdef}
\end{eqnarray}
for every allowed value of the involved parameters. Notice that the corresponding postmeasurement states $|\hat{n}_1\rangle$ and $|\hat{n}_2\rangle$ differ from the states $|1\rangle$ and $|0\rangle$.

According to Eq. (\ref{Absrho01shortNNselectLINEAR}), the magnitude of the coherence term decreases, or increases, with approximately constant velocity $\left|\mathfrak{V}\right|$, if $\mathfrak{V}/\omega_s<0$, or $\mathfrak{V}/\omega_s>0$, respectively, over short times, $t \ll t_l$. This property allows to determine the NMSs $\mathfrak{S}_{\rm d}$ which provide the short-time decreasing behavior, and the NMSs $\mathfrak{S}_{\rm i}$ which induce the short-time increasing behavior. These special schemes are described in the Appendix for the sake of fluency. In case the parameter $\mathfrak{V}$ vanishes, the magnitude of the coherence term increases (decreases) approximately quadratically over short times, $t\ll t_s$, if $a_1^2 >_{\left(<\right)}1+\eta_{0,T}\eta_{-1,0}^{-2}$. Again, the appearance of the short-time increasing or decreasing behavior is determined by the NMS, the temperature $T$ and the SD. 

\begin{figure}[t]
\centering
\includegraphics[height=4.75 cm, width=7.75 cm]{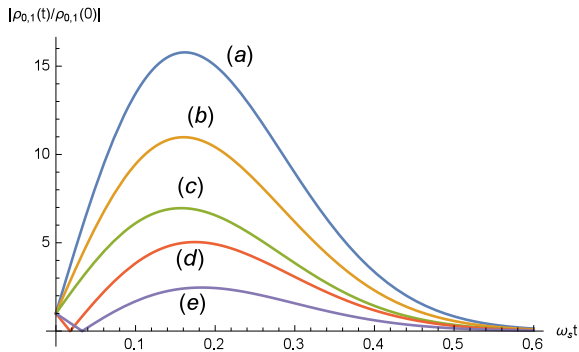}
\vspace*{0cm}
\caption{(Color online) Quantity $\left|\rho_{1,0}(t)/ \rho_{1,0}(0)\right|$ versus $\omega_s t$ 
for $0\leq \omega_s t \leq 0.6$, $J(\omega)=  \omega_s \left(\omega/\omega_s\right)^{\alpha}\exp\left(- \omega/\omega_s\right)$, $\alpha=1/2$, $\omega_s/T=1/10$, $\omega_0/T=1/100$, $\vartheta_0=\pi/8$, $q=Q\left(1/100,\pi/8\right)$ and different values of the angle $\Delta_{\zeta}$. Curve $(a)$ corresponds to $\Delta_{\zeta}=3.12$; $(b)$ to $\Delta_{\zeta}=3.11$; $(c)$ $\Delta_{\zeta}=3.09$; $(d)$ to $\Delta_{\zeta}=3.20$; $(e)$ to $\Delta_{\zeta}=3.25$.}
\label{Fig1}
\end{figure}

\begin{figure}[t]
\centering
\includegraphics[height=4.75 cm, width=7.75 cm]{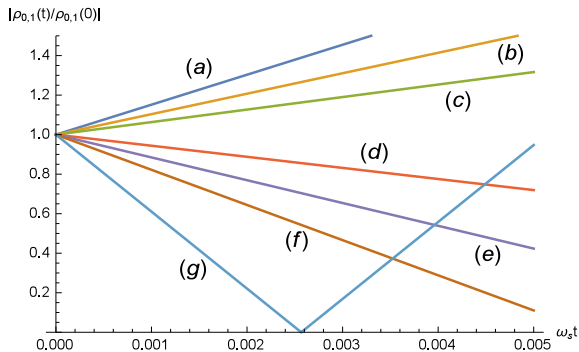}
\vspace*{0cm}
\caption{(Color online) Quantity $\left|\rho_{1,0}(t)/ \rho_{1,0}(0)\right|$ versus $\omega_s t$ 
for $0\leq \omega_s t \leq 0.005$, $J(\omega)=  \omega_s \left(\omega/\omega_s\right)^{\alpha}\exp\left(- \omega/\omega_s\right)$, $\alpha=1/2$, $\omega_s/T=1/10$, $\omega_0/T=1/100$, $\vartheta_0=\pi/8$, $q=Q\left(1/100,\pi/8\right)$ and different values of the angle $\Delta_{\zeta}$. Curve $(a)$ corresponds to $\Delta_{\zeta}=3.12$; $(b)$ to $\Delta_{\zeta}=3.11$; $(c)$ $\Delta_{\zeta}=3.09$; $(d)$ to $\Delta_{\zeta}=3.20$; $(e)$ to $\Delta_{\zeta}=3.17$, $(f)$ to $\Delta_{\zeta}=3.16$; $(g)$ to $\Delta_{\zeta}=3.15$.}
\label{Fig2}
\end{figure}

The time evolution of the magnitude of the coherence term of the reduced density matrix of the qubit is displayed in Figures \ref{Fig1} and, over short times, in Figure \ref{Fig2}, in case various NPMs are adopted. The corresponding NMSs differ from any of the special NMSs $\mathfrak{S}_0$, and coherence does not identically vanish during the whole reduced evolution. In Figure \ref{Fig1}, curves (a), (b) and (c) show an increasing behavior of the magnitude of the coherence term over short times; while curves (d) and (e) display, initially, decreasing 
coherence. Over intermediate times, coherence increases up to the corresponding maximum value,  and tends to vanish over long times. In Figure \ref{Fig2}, short-time linear increasing behaviors of coherence is shown by curves (a), (b) and (c); while curves (d), (e), (f) and (g) exhibit, initially, linear decreasing coherence.

\section{Maximizing the initial postmeasurement coherence 
}\label{4}

In the present dephasing model, the preparation of the initial state of a qubit with nonselective measurement induces limitations on the initial postmeasurement value of coherence,
\begin{eqnarray}
\left|\rho_{0,1}(0)\right|\leq\frac{1}{2}. \label{rho0half}
\end{eqnarray}
The upper bound, $\left|\rho_{0,1}(0)\right|=1/2$, is obtained by adopting any of the special NMSs $\mathfrak{S}_{\rm M}$, which are defined by the following relations:
\begin{eqnarray}
\hspace{-2em} 
\vartheta_1=\vartheta_2 =\frac{\pi}{2}, \hspace{1em}\Delta_{\zeta}=0, \hspace{1em}\forall \,\, \vartheta_0\in 
\left[0,\pi\right], \hspace{1em}\forall \,\,\zeta_0 \in \left[\right. 0,2\pi\left[
\right. . \label{SNMSRho12}
\end{eqnarray}
The postmeasurement states of these special schemes coincide one with another, $$|\hat{n}_1\rangle=|\hat{n}_2\rangle=2^{-1/2}\left[|1\rangle+ \exp \left(\imath \zeta_1 \right)|0\rangle\right].$$ 
The NMSs with common postmeasurement states, $|\hat{n}_1\rangle=|\hat{n}_2\rangle$, are described via the corresponding effects in Ref. \cite{M0}. Notice that the maximum value $1/2$ is realized for every allowed value of the polar angle $\vartheta_0$, of the azimuth angle $\zeta_0$, and of the common azimuth angles $\zeta_1$ and $\zeta_2$. 

By adopting any of the special NPMs $\mathfrak{S}_{\rm M}$, the magnitude of the coherence term decreases from the maximum value $1/2$ approximately quadratically over short times,
\begin{eqnarray}
\left|\rho_{0,1}(t)\right|\sim \frac{1}{2}+
 \mathfrak{W}^{\prime}t^2,
\label{Absrho01shortNNselectNMS1}
\end{eqnarray}
 for $t \ll t_{\rm s}$. The coefficient $\mathfrak{W}^{\prime}$, defined in the Appendix, is negative, $\mathfrak{W}^{\prime}/\omega_s^2<0$, and is approximated by the following form: $\mathfrak{W}^{\prime}\simeq -\eta_{0,0} / 4$, in the low-temperature limit, $T \ll \omega_0$. Therefore, by adopting any of the special NMSs $\mathfrak{S}_{\rm M}$, coherence decrease as follows in the low-temperature limit:
\begin{eqnarray}
\left|\rho_{0,1}(t)\right|\sim \frac{1}{2}-\frac{\eta_{0,0}}{ 4}t^2,
\label{Absrho01shortNNselectNMS1T0}
\end{eqnarray}
 for $T\ll \omega_0$. Notice that the coefficient of the negative quadratic term, $\eta_{0,0}/4$, is determined uniquely by the integral properties of the SD.

The upper bound $1/2$ of the initial postmeasurement coherence term is realized in the low-temperature limit, $T \ll \omega_0 $,
\begin{eqnarray}
&&\hspace{-1em}\lim_{\left(T/\omega_0\right)\to 0^+}\left|\rho_{1,0}(0)\right|=\left(\frac{1}{2}\right)^-,\label{rho010T012}
\end{eqnarray}
if, instead of the NMSs $\mathfrak{S}_{\rm M}$, the initial state of the qubit is prepared with further special NMSs. These special schemes are labeled, here, as $\mathfrak{S}_{\rm M}^{\prime}$ and are defined in the Appendix. By preparing the initial state of the qubit with any of the special NMSs $\mathfrak{S}_{\rm M}^{\prime}$, the short-time evolution of coherence is properly described by Eq. (\ref{Absrho01shortNNselectNMS1}). In the high-temperature limit, $T\gg \omega_0$, the magnitude of the coherence term approaches the maximum value $1/2$ uniquely in case any of the special NPMs $\mathfrak{S}_{\rm M}$ are adopted. This behavior is expected as these special schemes provide maximal coherence for every value of the temperature $T$. Differently from the low-temperature limit, no NPM provides the upper bound $1/2$ of coherence in the high-temperature limit, besides the special NPMs $\mathfrak{S}_{\rm M}$.

\begin{figure}[t]
\centering
\includegraphics[height=4.75 cm, width=7.75 cm]{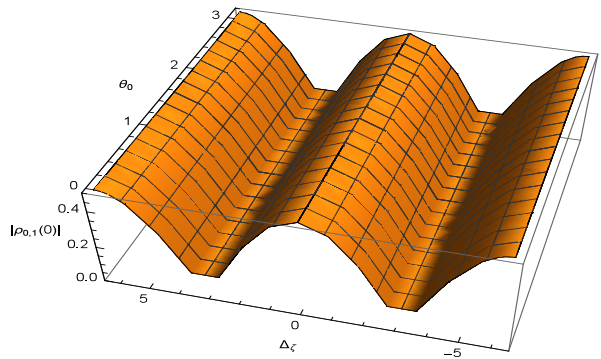}
\vspace*{0cm}
\caption{(Color online) Initial magnitude of the postmeasurement coherence term,
$ \left|\rho_{1,0}(0)\right|$, for $\omega_0/T=1/1000$, $\vartheta_1=\vartheta_2=\pi/2$, $0\leq \vartheta_0\leq \pi$ and $\left(-2\pi\right) <\Delta_{\zeta}<2 \pi$.}
\label{Fig3}
\end{figure}

\begin{figure}[t]
\centering
\includegraphics[height=4.75 cm, width=7.75 cm]{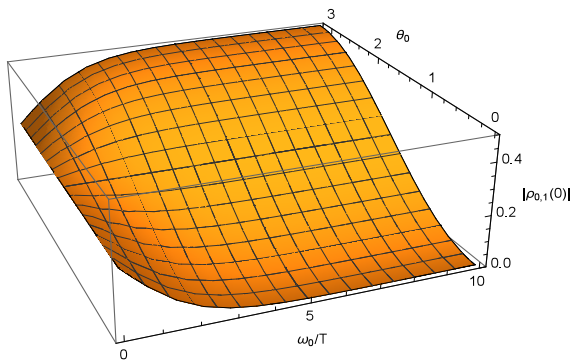}
\vspace*{0cm}
\caption{(Color online) Initial magnitude of the postmeasurement coherence term,
$ \left|\rho_{1,0}(0)\right|$, for $\vartheta_1=\pi/2$, $\vartheta_2=0,\pi$, $\Delta_{\zeta}=\pi/4$, $0<\omega_0/T\leq 10$ and $0\leq \vartheta_0\leq \pi$.}
\label{Fig4}
\end{figure}

The magnitude of the initial postmeasurement coherence term of the reduced density matrix of the qubit is displayed in Figure \ref{Fig3}, for various NMSs, and in Figure \ref{Fig4}, for various values of the polar angle $\vartheta_0$ and of the ratio $\omega_0/T$. In Figure \ref{Fig3}, the maximum value $1/2$ is obtained uniquely for $\Delta_{\zeta}=0$. The NMSs which provide maximal coherence reproduce the special NMSs $\mathfrak{S}_{\rm M}$ uniquely in this case. In Figure \ref{Fig4}, the maximum value $1/2$ is realized uniquely for $\vartheta_0=\pi$ and in the low-temperature limit, $T\ll \omega_0$. The NMSs reproduce a particular case of the special NMSs $\mathfrak{S}_{\rm M}^{\prime}$ uniquely in this condition.

\section{Maximizing the short-time increase or decrease of coherence 
}\label{5}

 In Section \ref{3}, it is described how the NPM determines the linear increasing or decreasing behavior of the magnitude of the coherence term, over short times. In the previous Section, it is shown how the measurement apparatus determines, bounds and maximizes the magnitude of the initial postmeasurement coherence via the features of the NMS. Similarly, in the present Section, it is described how the NPM determines, bounds and maximizes the velocity with which the magnitude of the coherence term increases or decreases over short times.

The magnitude $\left|\rho_{0,1}(t)\right|$ evolves algebraically over short times according to Eq. (\ref{Absrho01shortNNselect}). This magnitude varies with approximately constant velocity $\mathfrak{V}$, for $t\ll t_{\rm l}$, in case $\mathfrak{V}\neq 0$. If we consider the value of the velocity $\mathfrak{V}$ for every configuration of the postmeasurement states $|\hat{n}_1\rangle$ and $|\hat{n}_2\rangle$, we find that this velocity exhibits the maximum value $\mathfrak{V}_M\left(\vartheta_0, \omega_0/T\right)$ and the minimum value $\mathfrak{V}_m\left(\vartheta_0, \omega_0/T\right)$. The maximum value depends on the polar angle $\vartheta_0$, which characterizes the state $|\hat{n}_0\rangle$, and the ratio $\omega_0/T$ as follows,
\begin{eqnarray}
&&\hspace{-2.4em}\mathfrak{V}_M\left(\vartheta_0, \frac{\omega_0}{T}\right)=
2 \eta_{-1,0}\exp\left(-\frac{\omega_0}{T}\right)\left|\cos \vartheta_0\right|
\left[1+\exp\left(- \frac{\omega_0}{T}\right)\right]^{-1}\nonumber \\ 
&&\hspace{4.8em}\times\left\{1+\exp\left(- \frac{\omega_0}{T}\right)
+ \left|\cos \vartheta_0\right|\left[1-\exp\left(-\frac{\omega_0}{T}\right)\right]\right\}^{-1}. \label{VM}
\end{eqnarray}
The regime of maximal velocity $\mathfrak{V}_M\left(\vartheta_0,\omega_0/T\right)$ of the short-time increasing coherence is induced by special NMSs which are labeled, here, as $\mathfrak{S}_{\rm MV}$ and are described in the Appendix. This regime is approached in case the NMS tends to any of the special NMSs $\mathfrak{S}_0$ in the following ways:
\begin{eqnarray}
&&\hspace{-3em} 
q=Q\left(\frac{\omega_0}{T}, \vartheta_0\right), \hspace{1em}
 \Delta_{\zeta}\to \pm \pi ^{-}, \hspace{1em}\forall \,\, \vartheta_0\in 
\left[0,\frac{\pi}{2}\right.\Big[, \hspace{1em}\forall \,\,\zeta_0 \in \left[\right. 0,2\pi\left[\right. , \label{SNMSmaxQ1}\\
&&\hspace{-3em} 
q=Q\left(\frac{\omega_0}{T}, \vartheta_0\right), \hspace{1em}
 \Delta_{\zeta}\to \pm \pi ^{+}, \hspace{1em}\forall \,\, \vartheta_0\in \Big]\frac{\pi}{2}, \pi\Big], \hspace{1em}\forall \,\,\zeta_0 \in \left[\right. 0,2\pi\left[\right. ,\label{SNMSmaxQ2}  
\end{eqnarray}
with $\Delta_{\zeta}\neq \pm \pi$.

The minimum value $\mathfrak{V}_m\left(\vartheta_0, \omega_0/T\right)$ of the velocity $\mathfrak{V}$ is the opposite of the maximum value, 
\begin{eqnarray}
\hspace{-2.4em}
\mathfrak{V}_m\left(\vartheta_0, \frac{\omega_0}{T}\right)=-
\mathfrak{V}_M\left(\vartheta_0, \frac{\omega_0}{T}\right), \label{Vm}
\end{eqnarray}
for every allowed value of the involved parameters. The regime of minimal velocity $\mathfrak{V}_m\left(\vartheta_0,\omega_0/T\right)$ corresponds to the condition of maximally decreasing short-time coherence. This condition is induced by special NMSs which are labeled, here, as $\mathfrak{S}_{\rm mV}$, and are described in the Appendix. Again, this regime is approached in case the NMS tends to any of the special NMSs $\mathfrak{S}_0$ in the following ways:
\begin{eqnarray}
&&\hspace{-3em} 
q=Q\left(\frac{\omega_0}{T}, \vartheta_0\right), \hspace{1em}
 \Delta_{\zeta}\to \pm \pi ^{+}, \hspace{1em}\forall \,\, \vartheta_0\in 
\left[0,\frac{\pi}{2}\right.\Big[, \hspace{1em}\forall \,\,\zeta_0 \in \left[\right. 0,2\pi\left[\right., \label{SNMSminQ1}\\
&&\hspace{-3em} 
q=Q\left(\frac{\omega_0}{T}, \vartheta_0\right), \hspace{1em}
 \Delta_{\zeta}\to \pm \pi ^{-}, \hspace{1em}\forall \,\, \vartheta_0\in \Big]\frac{\pi}{2}, \pi\Big], \hspace{1em}\forall \,\,\zeta_0 \in \left[\right. 0,2\pi\left[\right. ,\label{SNMSminQ2}  \end{eqnarray}
with $\Delta_{\zeta} \neq \pm \pi$.

At this stage, it is useful to compare the NMSs (\ref{FDNMS1}) and (\ref{SNMSmaxQ1}), (\ref{SNMSmaxQ2}), or (\ref{SNMSminQ1}), (\ref{SNMSminQ2}). The special NMSs $\mathfrak{S}_0$, given by relations (\ref{FDNMS1}), are characterized by the condition $\Delta_{\zeta}= \pm \pi$, and provide full decoherence in the whole reduced dynamics of the qubit. Instead, the NMSs (\ref{SNMSmaxQ1}) are characterized by the following limiting property: $\Delta_{\zeta}\to \pm \pi^-$, for every $\vartheta_0 \in \big[0,\pi/2\big[$. Similarly, the NMSs (\ref{SNMSmaxQ2}) are characterized by the following limiting property: $\Delta_{\zeta}\to \pm \pi^+$, for every $\vartheta_0 \in \big]\pi/2,\pi\big]$. These conditions provide the regime of maximal velocity, $\mathfrak{V}_M\left(\vartheta_0, \omega_0/T\right)$, with which the magnitude of the coherence term increases over short times. Instead, the NMSs (\ref{SNMSminQ1}) are characterized by the limiting property $\Delta_{\zeta}\to \pm \pi^+$, for every $\vartheta_0 \in \big[0,\pi/2\big[$. Similarly, the NMSs (\ref{SNMSminQ2}) are characterized by the limiting property $\Delta_{\zeta}\to \pm \pi^-$, for every $\vartheta_0 \in \big]\pi/2,\pi\big]$. These conditions provide the regime of maximal decreasing velocity, $\left|\mathfrak{V}_m\left(\vartheta_0, \omega_0/T\right)\right|$, of the magnitude of the coherence term at short times. Hence, discontinuous transitions from full decoherence to the maximally increasing regime, or to the maximally decreasing regime, appear in the short-time evolution of the magnitude of the coherence term, $\left|\rho_{1,0}(t)\right|$, in case any of the above-mentioned NMSs departs from each of the conditions $\Delta_{\zeta}= \pm \pi$, as follows: $\Delta_{\zeta}\to \pm \pi^-$, or $\Delta_{\zeta}\to \pm \pi^+$. 

The maximal velocity $\mathfrak{V}_M\left(\vartheta_0, \omega_0/T\right)$ exhibits the following bounds by varying the involved parameters,
\begin{eqnarray}
0\leq\mathfrak{V}_M\left(\vartheta_0, \frac{\omega_0}{T}\right)
<\frac{ \eta_{-1,0}}{2}. \label{boundsV}
\end{eqnarray}
The low-temperature limit, $T \ll \omega_0$, or the condition $\vartheta_0=\pi/2$, provides the vanishing lower bound. Instead, the maximal velocity $\mathfrak{V}_M\left(\vartheta_0,\omega_0/T\right)$ exhibits the following high-temperature limit, 
\begin{eqnarray}
\lim_{\left(\omega_0/T\right)\to 0^+} \mathfrak{V}_M\left(\vartheta_0, \frac{\omega_0}{T}\right)= \frac{\eta_{-1,0}}{2}\left|\cos \vartheta_0\right|.
\label{VlT}
\end{eqnarray} 
Consequently, the maximal velocity $\mathfrak{V}_M\left(\vartheta_0, \omega_0/T\right)$ tends to the supremum value $ \eta_{-1,0}/2$ in the high-temperature limit, $T \gg \omega_0 $, if the initial condition is provided by the NMSs $\mathfrak{S}_{\rm MV}$ and (\ref{SNMSmaxQ1}), with $\vartheta_0=0$, or the NMSs $\mathfrak{S}_{\rm MV}$ and  (\ref{SNMSmaxQ2}), with $\vartheta_0=\pi$. Similarly, the magnitude $\left|\mathfrak{V}_m\left(\vartheta_0, \omega_0/T\right)\right|$ of the minimal velocity tends to the upper bound $\eta_{-1,0}/2$ in the high-temperature limit, $T \gg \omega_0 $, if the initial condition is realized by the NMSs $\mathfrak{S}_{\rm mV}$ and (\ref{SNMSminQ1}), with $\vartheta_0=0$, or the NMSs $\mathfrak{S}_{\rm mV}$ and (\ref{SNMSminQ2}), with $\vartheta_0=\pi$. These conditions provide the infimum value $\left(-\eta_{-1,0}/2\right)$ of the minimal velocity $\mathfrak{V}_m\left(\vartheta_0, 
\omega_0/T\right)$. Notice that these supremum and infimum values are determined uniquely by the first negative moment $\eta_{-1,0}$ of the SD.

The special NMSs $\mathfrak{S}_{\rm MV}$ and $\mathfrak{S}_{\rm mV}$ provide the same magnitude of the initial postmeasurement coherence, which is given by the expression below,
\begin{eqnarray}
&&\left|\rho_{0,1}(0)\right|=
\left\{1+\exp\left(-\frac{\omega_0}{T}\right)-
\left[1-\exp\left(- \frac{\omega_0}{T}\right)\right]
\left|\cos \vartheta_0\right|\right\}
\left|\tan \Delta_{\zeta}\right|\nonumber \\ &&\hspace{4.8em}\times
\left\{4\left[1+\exp\left(- \frac{\omega_0}{T}\right)\right]\right\}^{-1}.
   \label{R100SNMSMaxIncr12}
   \end{eqnarray}
By adopting any of these special NPMs and varying the angle $\Delta_{\zeta}$, the maximum value of expression (\ref{R100SNMSMaxIncr12}) is given by the following form:
\begin{eqnarray}
&&\hspace{-2em}\left|\rho_{0,1}(0)\right|=
\left\{\left[1-\exp\left(-\frac{\omega_0}{T}\right)\right]
\left|\cos \vartheta_0\right|\right\}^{1/2}
\left\{4\left[1+\exp\left(-\frac{\omega_0}{T}\right)\right]\right\}^{-1/2}.
   \label{R100SNMSmaxQ12}
\end{eqnarray}
This amount of initial postmeasurement coherence is realized by any of the special NMSs $\mathfrak{S}_{\rm MV}^{\prime}$, which are defined in the Appendix. Expression (\ref{R100SNMSmaxQ12}) tends to the value $1/2$ in the low-temperature limit, $T\ll \omega_0$, if $\left|\cos \vartheta_0\right|=1$. Hence, the  maximum value of the initial postmeasurement coherence is approached by adopting any of the special NPMs $\mathfrak{S}_{\rm MV}^{\prime}$ with $\vartheta_0=0$ or $\vartheta_0=\pi$, in the low-temperature limit, $T\ll \omega_0$. Instead, expression (\ref{R100SNMSmaxQ12}) vanishes in the high-temperature limit, $T \gg \omega_0$. Also, any of the NMSs (\ref{SNMSmaxQ1}), (\ref{SNMSmaxQ2}), (\ref{SNMSminQ1}) and (\ref{SNMSminQ2}) provide vanishing value of the initial postmeasurement coherence.

In summary, the analysis which is performed in the present and previous Section allows to compare the conditions which provide maximal magnitude of the initial postmeasurement coherence and maximal short-time increase of coherence. As matter of fact, the magnitude of the initial postmeasurement coherence is maximized by the value $1/2$. This value is realized by preparing the initial state of the qubit with any of the special NMSs $\mathfrak{S}_{\rm M}$. The maximum value $1/2$ is obtained in the low-temperature limit, $T \ll \omega_0$, by adopting the special NMSs $\mathfrak{S}_{\rm M}^{\prime}$. Instead, the  velocity $\mathfrak{V}$ tends to the value $ \eta_{-1,0}/2$, in the high-temperature limit, $T \gg \omega_0 $, in case further special NPMs are adopted. Furthermore, the initial magnitude of the postmeasurement coherence vanishes in the high-temperature limit. These observations suggest that the conditions which maximize the short-time increase of quantum coherence do not favor the maximization of initial post-measurement coherence, which is induced by further special NPMs or approached in the low-temperature limit.

\begin{figure}[t]
\centering
\includegraphics[height=4.75 cm, width=7.75 cm]{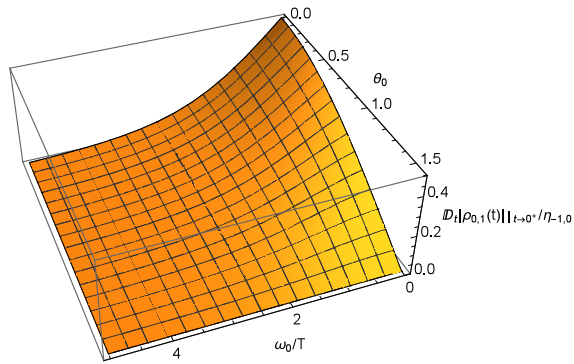}
\vspace*{0cm}
\caption{(Color online) Quantity $D_t \left|\rho_{0,1}(t)\right|\Big|_{t\to 0^+}\Big/\eta_{-1,0}$ for $0<\omega_0/T\leq 5$ and $0\leq \vartheta_0< \pi/2$, in case the initial state of the qubit is 
prepared with the special NMSs $\mathfrak{S}_{\rm MV}$.}
\label{Fig5}
\end{figure}

\begin{figure}[t]
\centering
\includegraphics[height=4.75 cm, width=7.75 cm]{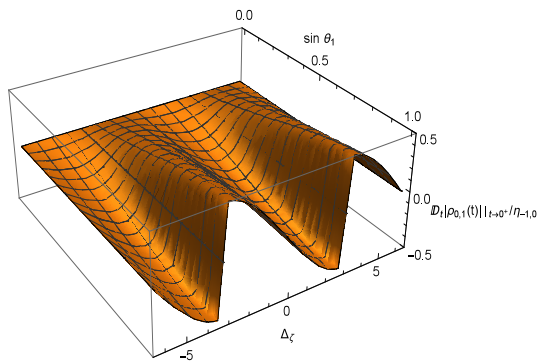}
\vspace*{0cm}
\caption{(Color online) Quantity $D_t \left|\rho_{0,1}(t)\right|\Big|_{t\to 0^+}\Big/\eta_{-1,0}$ for $\omega_0/T=1/10000$, $\vartheta_0=9\pi/10$, $\vartheta_2=\pi/2$, $0\leq  \sin \vartheta_1\leq 1$ and $\left(-2\pi\right) < \Delta_{\zeta}<2\pi$. The special vale of the parameter $q$ is $Q\left(1/10000,9\pi/10\right)\simeq 1$ and $q= \sin \vartheta_1$ for every $\vartheta_1\in \left[0,\pi\right]$.}
\label{Fig6}
\end{figure}

\begin{figure}[t]
\centering
\includegraphics[height=4.75 cm, width=7.75 cm]{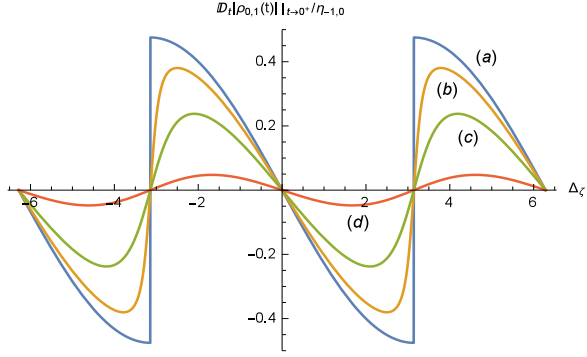}
\vspace*{0cm}
\caption{(Color online) Quantity $D_t \left|\rho_{0,1}(t)\right|\Big|_{t\to 0^+}\Big/\eta_{-1,0}$ for $\omega_0/T=1/10000$, $\vartheta_0=9\pi/10$, $\vartheta_2=\pi/2$, 
$\left(-2\pi\right) < \Delta_{\zeta}<2\pi$, and different values of the parameter $q$. Curve $(a)$ corresponds to $ q=Q\left(1/10000,9\pi/10\right)$; $(b)$ to $ q= 4/5$; $(c)$ to $q= 1/2$; $(d)$ to $ q= 1/10$.}
\label{Fig7}
\end{figure}

The velocity with which the magnitude of the coherence term evolves initially is displayed in Figure \ref{Fig5}, by considering the special NMSs $\mathfrak{S}_{\rm MV}$ and various values of the polar angle $\vartheta_0$ and of the ratio $\omega_0/T$. This velocity is displayed in Figure \ref{Fig6} and  Figure \ref{Fig7} at low temperature by considering various NMSs which reproduce condition (\ref{SNMSmaxQ2}). In Figure \ref{Fig5}, the supremum value of the short-time velocity is approached in the high-temperature limit for $\vartheta_0=0$. Figures \ref{Fig6} and \ref{Fig7} show that a discontinuous transition from the maximally decreasing regime of coherence to vanishing coherence, and to the maximally increasing regime occurs, by increasing the angle $\Delta_{\zeta}$ around the values $\Delta_{\zeta}=\pm \pi$, uniquely if the parameter $q$ is equal to the corresponding critical value. Particularly, in Figure \ref{Fig7}, curve (a) displays the discontinuous transition from the minimum value of the initial velocity, to the vanishing value, and to the maximum value by increasing the angle $\Delta_{\zeta}$ around the values $\Delta_{\zeta}=\pm \pi$. This behavior reproduces a particular case of conditions (\ref{SNMSminQ2}) and (\ref{SNMSmaxQ2}), respectively.

\section{Summary and conclusions}\label{6}

In a purely dephasing model of a qubit, the initial-state NPMs bound the initial amount of coherence. The maximal value $1/2$ of the initial postmeasurement coherence is realized from the thermal state of the whole system by performing special NPMs on the qubit. The same maximal value is obtained in the low-temperature limit by adopting further special NMSs. Hence, the choice of special NPMs and the low temperature of the initial thermal state of the whole system favor the maximization of the initial coherence of the qubit. In these conditions, the maximal magnitude of the coherence term decreases approximately quadratically over short times.

The measurement apparatus determines, over short times, an approximately linear increase or decrease of coherence via the NPM. The maximum value of the short-time increasing or decreasing velocity, which is obtained by considering every configuration of the postmeasurement states, is realized by special NMSs. The supremum value of the maximal increasing or decreasing velocity is obtained, under special conditions, in the high-temperature limit. Full decoherence of a qubit is induced over the whole reduced evolution by further special NPMs. If the NMS continuously departs from any of these special schemes in proper ways, maximal increase or maximal decrease of coherence appears over short times. Hence, a discontinuous transition from the regime of full decoherence to the regime of maximally increasing or maximally decreasing coherence is induced, over short times, by properly changing the NMS. 

In conclusion, in the purely dephasing model of a qubit, the NPM influences relevantly the initial postmeasurement value and the short-time evolution of quantum coherence. These properties might be of interest to manipulate and develop control of quantum coherence of a qubit via the setting of the measurement apparatus \cite{M0,measure7,BP,measure2}.

\appendix
\section{Details}\label{A}
 
The magnitude of the initial postmeasurement coherence term, $\left|\rho_{0,1}(0)\right|$, is evaluated in straightforward way from Eq. (\ref{rho01t0NNselect}). The maximum value, which is obtained by considering every configuration of the postmeasurement states, is studied by analyzing the quantity $\left|\rho_{0,1}(0)\right|^2$ as a function of the three variables $u$, $v$ and $\Delta_{\zeta}$, in the closed domain $\bar{\mathbb{D}}$. This function is labeled, here, as $\mathfrak{R}\left(u,v,\Delta_{\zeta}\right)$. The variables  $u$ and $v$ are defined as follows: $u\equiv \sin \vartheta_1$ and $v\equiv \sin \vartheta_2$, while the set $\mathbb{D}$ is given by the following form: $\mathbb{D}=\left[0,1\right]\times \left[0,1\right]\times \left.\right]-2 \pi, 2 \pi \left[\right.$. The polar angle $\vartheta_0$, the azimuth angle $\zeta_0$, the transition frequency $\omega_0$ and the temperature $T$ are treated as parameters. The stationary points of the function $\mathfrak{R}\left(u,v,\Delta_{\zeta}\right)$ are the solutions of the following system of three equations: 
$\partial_u \mathfrak{R}\left(u,v,\Delta_{\zeta}\right) =0$, 
$\partial_v \mathfrak{R}\left(u,v,\Delta_{\zeta}\right) =0$, 
$\partial_{\Delta_{\zeta}}\mathfrak{R}\left(u,v,\Delta_{\zeta}\right) =0$. This system of equations is equivalent to the following one: 
\begin{eqnarray}
&&\hspace{-2em} u \left[1+ w-\left(w-1\right) \cos \vartheta_0\right]
+v \left[1+ w+\left(w-1\right) \cos \vartheta_0\right]
 \cos \Delta_{\zeta}=0, \label{gradEqu}\\
&&\hspace{-2em} u \left[1+ w-\left(w-1\right) \cos \vartheta_0\right]
\cos \Delta_{\zeta}
+v \left[1+ w+\left(w-1\right) \cos \vartheta_0\right]
 =0, \label{gradEqv}\\
&&\hspace{-2em}u v \sin \Delta_{\zeta}=0, \label{gradEqd}
\end{eqnarray}
where $w\equiv\exp\left(\omega_0/T\right)$. The following vanishing solution: $u=v=0$, holds for every $\Delta_{\zeta}\in\left[-2\pi,2\pi\right]$ and for every allowed value of the involved parameters. The solution below, 
 \begin{eqnarray}
u=Q\left(\frac{\omega_0}{T},\vartheta_0
\right)
v, \hspace{1em} \Delta_{\zeta}=\pm \pi,\label{Sol1MR}
\end{eqnarray}
holds for every $ v \in \left[0,1/Q\left(\omega_0/T,\vartheta_0
\right)
\right]$, in case $\vartheta_0\in\left[0, \pi/2\right]$, and for every $ v \in \left[0, 1\right]$, in case $\vartheta_0\in\left[\pi/2,\pi\right]$. The function $\mathfrak{R}\left(u,v,\Delta_{\zeta}\right)$ vanishes for $u=v=0$ and at every point which is given by the condition (\ref{Sol1MR}). The function $\mathfrak{R}\left(u,v,\Delta_{\zeta}\right)$ is continuous in the closed and bounded domain $\bar{\mathbb{D}}$ and the vanishing global minimum occurs at every stationary point which belongs to the interior of this domain. Hence, the global maximum of the function $\mathfrak{R}\left(u,v,\Delta_{\zeta}\right)$ exists and occurs at the boundary of the domain $\bar{\mathbb{D}}$. This boundary is characterized by any of the following conditions: $u=0$, $u=1$, $v=0$, $v=1$, 
$\Delta_{\zeta}=-2\pi$, $\Delta_{\zeta}=2\pi$. The function $\mathfrak{R}\left(0,v,\Delta_{\zeta}\right)$, of the variables $v$ and $\Delta_{\zeta}$, exhibits the following maximum value:
\begin{eqnarray}
\frac{1}{16}\left(1 +\frac{w-1}{w+1} \cos \vartheta_0\right)^2,
\label{MR1}
\end{eqnarray}
 which occurs at $v=1$ for every $\Delta_{\zeta}\in \left[-2 \pi, 2 \pi \right]$. The function $\mathfrak{R}\left(u,0,\Delta_{\zeta}\right)$, of the variables $u$ and $\Delta_{\zeta}$, exhibits the following maximum value:
\begin{eqnarray}
\frac{1}{16}\left(1- \frac{w-1}{w+1} \cos \vartheta_0\right)^2,
\label{MR2}
\end{eqnarray}
which occurs at $u=1$ for every $\Delta_{\zeta}\in \left[-2 \pi, 2 \pi \right]$. The function $\mathfrak{R}\left(1,v,\Delta_{\zeta}\right)$, of the variables $v$ and $\Delta_{\zeta}$,
exhibits the maximum value $1/4$ which occurs at $v=1$ for $\Delta_{\zeta}=0, \pm 2\pi$. The function $\mathfrak{R}\left(u,1,\Delta_{\zeta}\right)$, of the variables $u$ and $\Delta_{\zeta}$, exhibits the maximum value $1/4$ which occurs at $u=1$ for $\Delta_{\zeta}=0, \pm 2\pi$. The function $\mathfrak{R}\left(u,v,\pm 2\pi\right)$, of the variables $u$ and $v$, exhibits the maximum value $1/4$ which occurs at $u=v=1$. The value $1/4$ is larger than any of the values (\ref{MR1}) and 
(\ref{MR2}) for every $\vartheta_0\in \left[0,\pi\right]$ and for every $w>0$. Hence, the value $1/2$ is the global maximum of the initial postmeasurement coherence $\left|\rho_{0,1}(0)\right|$ in the domain $\mathbb{D}$ and occurs at $u=v=1$ for $\Delta_{\zeta}=0$. This condition defines the special NMSs (\ref{SNMSRho12}).

The magnitude of the initial postmeasurement coherence term, 
Eq. (\ref{rho01t0NNselect}), is properly approximated by the expression below,
\begin{eqnarray}
&&\hspace{-0.4em}\left|\rho_{0,1}(0)\right|\simeq \frac{1}{2}
\left(\sin^4 \frac{\vartheta_0}{2}\sin^2 \vartheta_1+\cos^4 \frac{\vartheta_0}{2}\sin^2 \vartheta_2+\frac{1}{2}\sin^2 \vartheta_0
\sin \vartheta_1\sin \vartheta_2 \cos \Delta_{\zeta}
\right)^{1/2}, \nonumber \\ &&\hspace{-1em}\label{rho010T0}
\end{eqnarray}
 in the low-temperature limit, $T\ll \omega_0$. By adopting the above-reported procedure, we find that this expression exhibits the maximum value $1/2$ if the relations (\ref{SNMSRho12}) are fulfilled and, additionally, in case any of the following conditions hold:
\begin{eqnarray}
&&\hspace{-2.5em} 
\vartheta_0 = 0,\pi, \hspace{1em}\vartheta_1=\vartheta_2 =\frac{\pi}{2}, \hspace{1em}\forall \,\,\zeta_0 \in \left[\right. 0,2\pi\left[\right., \hspace{1em}\forall \,\, \Delta_{\zeta}\in 
\left.\right]-2\pi,2\pi\left[\right. , \label{SNMSRho12T00}
\\
&&\hspace{-2.5em} 
\vartheta_0 = 0, \hspace{1em}\vartheta_2 =\frac{\pi}{2}, \hspace{1em}\forall \,\,\zeta_0 \in \left[\right. 0,2\pi\left[\right.,
\hspace{1em}\forall \,\, \vartheta_1\in 
\left[0,\pi\right], \hspace{1em}\forall \,\, \Delta_{\zeta}\in 
\left.\right]-2\pi,2\pi\left[\right. ,\nonumber \\ && \label{SNMSRho12T01}
\\
&&\hspace{-2.5em} 
\vartheta_0 = \pi, \hspace{1em}\vartheta_1 =\frac{\pi}{2},\hspace{1em}\forall \,\,\zeta_0 \in \left[\right. 0,2\pi\left[\right.,
\hspace{1em}\forall \,\, \vartheta_2\in 
\left[0,\pi\right], \hspace{1em}\forall \,\, \Delta_{\zeta}\in 
\left.\right]-2\pi,2\pi\left[\right.. \nonumber \\ && \label{SNMSRho12T02}
\end{eqnarray}
These relations define the special NMSs $\mathfrak{S}_{\rm M}^{\prime}$. The condition (\ref{SNMSRho12T00}) with $\Delta_{\zeta}=0$, provides the maximal magnitude of the initial postmeasurement coherence at every temperature, but defines some of the special NMSs $\mathfrak{S}_{\rm M}$, and is, therefore, excluded from definition of the special NMSs $\mathfrak{S}_{\rm M}^{\prime}$. In the high-temperature limit, $T\gg \omega_0$, the magnitude of the postmeasurement coherence term is properly approximated by the expression below,
\begin{eqnarray}
&&\hspace{-1em}\left|\rho_{1,0}(0)\right|\simeq 
\frac{1}{4}\left( \sin^2 \vartheta_1+
\sin^2 \vartheta_2+2 \sin \vartheta_1  \sin \vartheta_2
 \cos \Delta_{\zeta} \right)^{1/2}, \label{rho010Tinfty}
\end{eqnarray}
for $T\gg \omega_0$. The maximum value $1/2$ occurs uniquely under the conditions which define the special NPMs $\mathfrak{S}_{\rm M}$.

 The short-time behavior of the coherence term is given by Eq. (\ref{Absrho01shortNNselect}) for $t \ll t_{\rm s}$. The coefficient $\mathfrak{V}$ of the linear term and the coefficient $\mathfrak{W}$ of the quadratic term are given by the expressions 
below \cite{G_IJQI2020}, 
\begin{eqnarray}
&&\mathfrak{V}=\left|\rho_{0,1}(0)\right|a_{2,T}\eta_{-1,0}, \label{Vdef}\\
&&\mathfrak{W}=\frac{1}{2}\left|\rho_{0,1}(0)\right|\xi_{T},\label{Wdef}
\end{eqnarray}
where 
\begin{eqnarray}
\xi_{T}
=\left(a^2_{1,T}-1\right)\eta_{-1,0}^2-\eta_{0,T} ,   \nonumber 
\end{eqnarray}
and
\begin{eqnarray}
\eta_{k,T}=\int_{\omega_g}^{\omega_M}\omega^k J_T\left(\omega\right) d \omega,  \nonumber 
\end{eqnarray} 
 for every $k=-1,\ldots,2$. The constant $a_{1,T}$, $a_{2,T}$ and $a_{3,T}$ are given by the following forms:
\begin{eqnarray}
&&a_{1,T}=\frac{N_{1,T}}{N_{0,T}}, \hspace{1em} a_{2,T}=\frac{N_2}{N_{0,T}}, \hspace{1em} a_{3,T}=\frac{N_{1,T}^2+N_{2}^2}{N_{0,T}^2}-1. \nonumber
\end{eqnarray}
The parameters $N_{0,T}$, $N_{1,T}$ and $N_{2}$ are defined in terms of the polar angles $\vartheta_0$, $\vartheta_1$, $\vartheta_2$, and of the azimuthal angles $\zeta_0$, $\zeta_1$, $\zeta_2$, of the unit vectors $\hat{n}_0$, $\hat{n}_1$, $\hat{n}_2$, respectively \cite{M0},
\begin{eqnarray}
&&\hspace{-3em}N_{0,T}=\left[\frac{1}{2}\sin^2\vartheta_0+
\exp\frac{\omega_0}{T}
\sin^4\frac{\vartheta_0}{2}+\exp\left(-\frac{\omega_0}{T}\right)
\cos^4\frac{\vartheta_0}{2}\right]\sin^2\vartheta_1 \nonumber \\&&\hspace{0.4em}
+\Bigg[\frac{1}{2}\sin^2\vartheta_0+
\exp\frac{\omega_0}{T}
\cos^4\frac{\vartheta_0}{2}+\exp\left(-\frac{\omega_0}{T}\right) \sin^4\frac{\vartheta_0}{2}\Bigg]\sin^2\vartheta_2   \nonumber \\&&\hspace{0.4em}
+\left[\cosh \frac{\omega_0}{T}\sin^2\vartheta_0+
2\left(\sin^4\frac{\vartheta_0}{2}+\cos^4\frac{\vartheta_0}{2}\right)\right]\cos \Delta_{\zeta}\sin\vartheta_1 \sin\vartheta_2,
\label{N0T}\\
&&\hspace{-3em}N_{1,T}=\left[\exp\frac{\omega_0}{T}\sin^4\frac{\vartheta_0}{2}-\exp\left(-\frac{\omega_0}{T}\right)\cos^4\frac{\vartheta_0}{2}\right]\sin^2\vartheta_1
\nonumber\\&&\hspace{0.4em}+\left[\exp\frac{\omega_0}{T}\cos^4\frac{\vartheta_0}{2}-\exp\left(-\frac{\omega_0}{T}\right)\sin^4\frac{\vartheta_0}{2}\right]\sin^2\vartheta_2
\nonumber \\&&\hspace{0.4em}
+\sinh \frac{\omega_0}{T} \sin^2\vartheta_0\cos\Delta_{\zeta}\sin\vartheta_1 \sin\vartheta_2,
\label{N1T}
\\
&&\hspace{-3em}N_{2}=2\cos\vartheta_0\sin \Delta_{\zeta}\sin\vartheta_1 \sin\vartheta_2. \label{N2}
\end{eqnarray}
The parameter $N_{0,T}$ fulfills the following relations for every allowed value of the involved parameters:
\begin{eqnarray}
N_{0,T}\geq A_0\left(\vartheta_0,\frac{\omega_0}{T}\right)
\left[
\sin \vartheta_1-Q\left(\frac{\omega_0}{T},\vartheta_0
\right)\sin \vartheta_2
\right]^2\geq 0, \label{N0geq0}
\end{eqnarray}
where
$$
A_0\left(\vartheta_0,\frac{\omega_0}{T}\right)=\frac{1}{2}\sin^2\vartheta_0+
\exp\frac{\omega_0}{T}
\sin^4\frac{\vartheta_0}{2}+\exp\left(-\frac{\omega_0}{T}\right)
\cos^4\frac{\vartheta_0}{2}.
$$
According to the relations (\ref{N0geq0}), the parameter $N_0$ is nonnegative, $N_0\geq 0$. Particularly, this parameter vanishes, $N_0=0$, if every postmeasurement state coincides with any of the state $|1\rangle$ and $|0\rangle$, 
\begin{eqnarray}
&&\hspace{-1em}\vartheta_1= 0,\pi,\hspace{0.4em}\vartheta_2= 0,\pi,\hspace{0.4em}\forall \,\,\vartheta_0\in \left[0,\pi\right], \hspace{0.4em}\forall \,\,\zeta_0 \in \left[\right. 0,2\pi\left[\right.,
 \hspace{0.4em}\forall \,\,\Delta_{\zeta} \in 0 \in \left.\right] -2\pi, 2 \pi\left[\right. .\nonumber \\ &&
\label{FDNMS0} 
\end{eqnarray}
 Additionally, the parameter $N_0$ vanishes under any of the conditions (\ref{FDNMS1}), which define the special NMSs $\mathfrak{S}_0$. These conditions provide vanishing coherence in the whole reduced evolution of the qubit, $\rho_{0,1}(t)=0$ for every $t \geq 0$, and, therefore, are excluded from the present analysis of the short-time regime of increasing or decreasing coherence. The short time scale, $t\ll t_{\rm s}$, is obtain by imposing the following three conditions: the short times are required to be small with respect to the typical time scale $1/\omega_s$ of the reduced evolution; the rest of the power series expansion of the dephasing factor $\Xi_T\left(t\right)$ and of the auxiliary function $\upsilon_0(t)$, around $t=0$, is requested to be small if compared to the second order truncation. In this way, Eq. (\ref{ts}) is obtained. The time $t_l$, given by Eq. (\ref{tl}), is derived in the same way, by requiring the same three conditions and, additionally, by considering Eq. (\ref{Absrho01shortNNselect}) and imposing the second order term, $\mathfrak{W} t^2$, to be small if compared to the first order term $\mathfrak{V} t$. In case the initial condition is prepared with any of the NPMs $\mathfrak{S}_{\rm M}$, Eq. (\ref{Absrho01shortNNselect}) provides the short-time evolution (\ref{Absrho01shortNNselectNMS1}) with $$ \mathfrak{W}^{\prime}=-\frac{\eta_{0,T}}{4}-\frac{\eta_{-1,0}^2\exp\left(-\omega_0/T\right)}{\left[1+\exp\left(-\omega_0/T\right)\right]^2}.$$

According to Eq. (\ref{Vdef}), the velocity $\mathfrak{V}$ is negative if the parameter $N_2$ is negative. This condition provides the relations below,
\begin{eqnarray}
&&\hspace{-0.7em}
\vartheta_1\neq 0,\pi,\hspace{0.3em}\vartheta_2\neq 0,\pi,\hspace{0.3em}
\forall \,\,\vartheta_0\in \Big[0,\frac{\pi}{2}\Big[, \hspace{0.3em}\forall \,\,\zeta_0 \in \left[\right. 0,2\pi\left[\right.,
\hspace{0.3em}
\forall \,\,\Delta_{\zeta}\in \left.\right]- \pi,0\left[\right. \cup
\left.\right]\pi,2\pi\left[\right.,\nonumber \\&&
\label{dephScheme1} \\
&&\hspace{-0.7em}
\vartheta_1\neq 0,\pi,\hspace{0.3em}\vartheta_2\neq 0,\pi,
\hspace{0.3em}
\forall \,\,\vartheta_0\in \Big]\frac{\pi}{2},\pi \Big], \hspace{0.3em}\forall \,\,\zeta_0 \in \left[\right. 0,2\pi\left[\right.,
\hspace{0.3em}
\forall \,\,\Delta_{\zeta}\in \left.\right]-2 \pi,-\pi\left[\right. \cup
\left.\right]0,\pi\left[\right.
. \nonumber \\&&\label{dephScheme2}
\end{eqnarray}
These relations define the special NMSs $\mathfrak{S}_{\rm d}$. The velocity $\mathfrak{V}$ is positive if the parameter $N_2$ is positive. This condition leads to the relations below, 
\begin{eqnarray}
&&\hspace{-0.7em}
\vartheta_1\neq 0,\pi,\hspace{0.2em}\vartheta_2\neq 0,\pi,\hspace{0.2em}
\forall \,\,\vartheta_0\in \Big[0,\frac{\pi}{2}\Big[, \hspace{0.2em}\forall \,\,\zeta_0 \in \left[\right. 0,2\pi\left[\right.,
\hspace{0.2em}
\forall \,\,\Delta_{\zeta}\in \left.\right]-2 \pi,-\pi\left[\right. \cup
\left.\right]0,\pi\left[\right., 
\nonumber \\&&
\label{inphScheme1} \\
&&\hspace{-0.7em}
\vartheta_1\neq 0,\pi,\hspace{0.2em}\vartheta_2\neq 0,\pi,
\hspace{0.2em}
\forall \,\,\vartheta_0\in \Big]\frac{\pi}{2},\pi\Big], 
\hspace{0.2em}\forall \,\,\zeta_0 \in \left[\right. 0,2\pi\left[\right.,
\hspace{0.2em}\forall \,\,
\Delta_{\zeta}\in \left.\right]- \pi,0\left[\right. \cup
\left.\right]\pi,2\pi\left[\right.. \nonumber \\&&\label{inphScheme2}
\end{eqnarray}
These relations define the special NMSs 
$\mathfrak{S}_{\rm i}$. According to Eq. (\ref{Vdef}), the velocity $\mathfrak{V}$ vanishes if the initial postmeasurement coherence vanishes, $\left|\rho_{0,1}(0)\right|=0$, or if the coefficient $a_{2,T}$ vanishes, $a_{2,T}=0$. Hence, this velocity vanishes if any of the following conditions holds: $\vartheta_0=\pi/2$; $\vartheta_1=0,\pi$; $\vartheta_2=0,\pi$; $\Delta_{\zeta}=0$; and $\Delta_{\zeta}=\pm\pi$ for $q \neq Q\left(\omega_0/T,\vartheta_0\right)$.

The maximum value $\mathfrak{V}_M$ and the minimum value $\mathfrak{V}_m$ of the velocity $\mathfrak{V}$, which are obtained by considering every configuration of the postmeasurement states, are found by studying the ratio $\mathfrak{V}/ \eta_{-1,0}$ as a function of the variables $u$, $v$ and $\Delta_{\zeta}$. This function is labeled, here, as $\mathfrak{U}\left(u,v,\Delta_{\zeta}\right)$ and is analyzed in the domain $\mathbb{D}$. The polar angle $\vartheta_0$, the azimuth angle $\zeta_0$, the frequency $\omega_0$ and the temperature $T$ are treated as parameters. By definition, the sign and the zeros of the function $\mathfrak{U}\left(u,v,\Delta_{\zeta}\right)$ coincide with the sign and the zeros of the velocity $\mathfrak{V}$, which are analyzed in the previous paragraph. The singular points of the function $\mathfrak{U}\left(u,v,\Delta_{\zeta}\right)$ are obtained from the following vanishing condition: $N_0=0$, and are given by the relations (\ref{Sol1MR}). These relations hold for every $v\leq1/Q \left(\vartheta_0,\omega_0/T\right)$, in case $\cos \vartheta_0\geq 0$, or for every $v\leq 1$ in case $\cos \vartheta_0\leq 0$, and provides the special NMSs $\mathfrak{S}_0$. The function $\mathfrak{U}\left(u,v,\Delta_{\zeta}\right)$ exhibits a discontinuous behavior in any of the singular points. In fact, this function vanishes in the limiting condition $\left(u/v\right)\to Q\left(\omega_0/T,\vartheta_0\right)$ for $\Delta_{\zeta}=\pm\pi$ and $\left(u/v\right)\neq Q\left(\omega_0/T,\vartheta_0\right)$, while the following limits hold in case $\Delta_{\zeta}\to \pm \pi$ for $u= Q\left(\omega_0/T,\vartheta_0\right)v$ and $\Delta_{\zeta}\neq\pm\pi$: 
\begin{eqnarray}
&& \hspace{-1.5em}\lim_{\Delta_{\zeta}\to \pm \pi^-}\mathfrak{U}\left( Q\left(\frac{\omega_0}{T},\vartheta_0\right)v,v,\Delta_{\zeta}\right)=
 -\lim_{\Delta_{\zeta}\to \pm \pi^+}\mathfrak{U}\left( Q\left(\frac{\omega_0}{T},\vartheta_0\right)v,v,\Delta_{\zeta}\right),
\label{VMexpr1}\\
&&\hspace{-1.5em} \lim_{\Delta_{\zeta}\to \pm \pi^+}\mathfrak{U}\left( Q\left(\frac{\omega_0}{T},\vartheta_0\right)v,v,\Delta_{\zeta}\right)=
- 2  v \exp \left(-\frac{\omega_0}{T}\right) \cos \vartheta_0  \Big\{\left[1+\exp \left(-\frac{\omega_0}{T}\right)\right]^2\nonumber \\ &&\hspace{14em}-\left[1-\exp \left(-2\frac{\omega_0}{T}\right)\right]\cos \vartheta_0\Big\}^{-1}.
\label{VMexpr2}
\end{eqnarray} 
These limits are obtained from the following relation:
\begin{eqnarray}
&&\hspace{-1.5em}  \mathfrak{U}\left( Q\left(\frac{\omega_0}{T},\vartheta_0\right)v,v,\Delta_{\zeta}\right)=
\pm 2  v \exp \left(-\frac{\omega_0}{T}\right) \cos \vartheta_0\sin \frac{\Delta_{\zeta}}{2}  \Big\{\left[1+\exp \left(-\frac{\omega_0}{T}\right)\right]^2\nonumber \\ &&\hspace{10.3em}-\left[1-\exp \left(-2\frac{\omega_0}{T}\right)\right]\cos \vartheta_0\Big\}^{-1},
\label{VMrel}
\end{eqnarray} 
which holds for $\cos \left(\Delta_{\zeta}/2\right)>_{\left(<\right)}0$, $\Delta_{\zeta}\neq \pi,-\pi$, and for every $v\leq1/Q \left(\vartheta_0,\omega_0/T\right)$, if $\cos \vartheta_0\geq 0$, or for every $v\leq 1$, if $\cos \vartheta_0\leq 0$.

The maximum and minimum values of the function $\mathfrak{U}\left(u,v,\Delta_{\zeta}\right)$ are investigated in the domain $\mathbb{D}\setminus \mathbb{S}$. The set $\mathbb{S}$ is composed by the elements $\left(u,v,\Delta_{\zeta}\right)$ of the domain such that the corresponding NMSs provide vanishing value of the initial postmeasurement coherence, $\rho_{0,1}\left(0\right)=0$. These elements are excluded as, in these cases, coherence identically vanishes during the whole reduced evolution. The set $\mathbb{S}$ is defined, for the sake of clarity, as the union of the sets $\mathbb{S}_0$ and $\mathbb{S}_1$, i.e., $\mathbb{S}=\mathbb{S}_0\cup\mathbb{S}_1$, where
\begin{eqnarray}
&&\hspace{-1.5em}  
\mathbb{S}_0= \left\{\left(u,v,\Delta_{\zeta}\right) \in \mathbb{D}\hspace{0.5em} |\hspace{0.5em}u=v=0\right\}, \nonumber \\
&&\hspace{-1.5em}  
\mathbb{S}_1= \left\{\left(u,v,\Delta_{\zeta}\right) \in \mathbb{D}\hspace{0.5em}|\hspace{0.5em}\frac{u}{v}=Q\left(\frac{\omega_0}{T},\vartheta_0\right), \hspace{0.5em}\Delta_{\zeta}=\pm\pi\right\}. \nonumber 
\end{eqnarray}
The set $\mathbb{S}_0$ corresponds to the NMSs (\ref{FDNMS0}). The set $\mathbb{S}_1$ is composed by the singular points of the function $\mathfrak{U}\left(u,v,\Delta_{\zeta}\right)$ and corresponds to the special NMSs $\mathfrak{S}_0$. The following partition of the domain $\mathbb{D}$ is performed for the sake of convenience: $\mathbb{D}=\bigcup_{j= -1}^2 \mathbb{D}_j \cup
\bigcup_{j= -1}^1 \mathbb{D}^{\left(\star\right)}_j,$ where $\mathbb{D}_j=\left[0,1\right] \times \left[0,1\right] 
\times\left.\right](j-1)\pi, j\pi\left[\right.$, 
for every $j=-1,\ldots,2$, and $\mathbb{D}^{\left(\star\right)}_{j}=\left[0,1\right] \times
\left[0,1\right] \times \left\{j \pi\right\}$, for every $j=-1,0,1$. According to the above-performed analysis, the function $\mathfrak{U}\left(u,v,\Delta_{\zeta}\right)$ is positive (negative) in the sets $\mathbb{D}_{-1}\setminus \mathbb{S}_0$ and $\mathbb{D}_{1}\setminus \mathbb{S}_0$ and negative (positive) in the sets $\mathbb{D}_{0}\setminus \mathbb{S}_0$ and $\mathbb{D}_{2}\setminus \mathbb{S}_0$ in case the term $\cos \vartheta_0$ is positive (negative), $\cos \vartheta_0>_{\left(<\right)}0$. The function 
$\mathfrak{U}\left(u,v,\Delta_{\zeta}\right)$ vanishes in every of the above-mentioned sets for $\vartheta_0=\pi/2$, and vanishes also in the sets $\mathbb{D}^{\left(\star\right)}_0$, $\mathbb{D}^{\left(\star\right)}_{-1}\setminus \mathbb{S}_1$ and $\mathbb{D}^{\left(\star\right)}_{1}\setminus \mathbb{S}_1$ for every $\vartheta_0\in \left[0,\pi\right]$.

The stationary points of the function $\mathfrak{U}\left(u,v,\Delta_{\zeta}\right)$ are solutions of the following equations: 
$\partial_u \mathfrak{U}\left(u,v,\Delta_{\zeta}\right) =0$, 
$\partial_v \mathfrak{U}\left(u,v,\Delta_{\zeta}\right) =0$, 
$\partial_{\Delta_{\zeta}} \mathfrak{U}\left(u,v,\Delta_{\zeta}\right)=0$. The conditions which provide the vanishing value of the function $\mathfrak{U}\left(u,v,\Delta_{\zeta}\right)$ are excluded. In this way, the first and second equations, $\partial_u \mathfrak{U} \left(u,v,\Delta_{\zeta}\right)=0$, and $\partial_v \mathfrak{U} \left(u,v,\Delta_{\zeta}\right)=0$, are equivalent to the following ones:
\begin{eqnarray}
&&\hspace{-2em} u \left[1+ w-\left(w-1\right) \cos \vartheta_0\right]\cos \Delta_{\zeta}+v \left[1+ w+\left(w-1\right)\cos \vartheta_0 \right]
 =0, \label{gradEqVu}\\
&&\hspace{-2em} u \left[1+ w-\left(w-1\right) \cos \vartheta_0\right]
+v \left[1+ w+\left(w-1\right) \cos \vartheta_0\right]\cos \Delta_{\zeta}
 =0. \label{gradEqVv}
\end{eqnarray}
The vanishing solution, $u=v=0$, of Eqs.(\ref{gradEqVu}) and (\ref{gradEqVv}) is unique except for $\Delta=0, \pm \pi$. Therefore, no stationary point of the function $\mathfrak{U}\left(u,v,\Delta_{\zeta}\right)$ exists in the interior of the set $\mathbb{D}_j$, for every $j=-1,\ldots,2$.

As far as the boundary values of the function $\mathfrak{U}\left(u,v,\Delta_{\zeta}\right)$, with $u=1$, are concerned, we find that the following equation: $\partial_v\mathfrak{U}\left(1,v,\Delta_{\zeta}\right)=0$, exhibits the solution below,
 \begin{eqnarray}
v=- \frac{\sec \Delta_{\zeta}}{Q\left(\omega_0/T,\vartheta_0
\right)}
,\label{condVM1}
\end{eqnarray}
for every allowed value of the involved parameters. This solution fulfills the following equation: $\partial_{\Delta_{\zeta}}\mathfrak{U}\left(1,v,\Delta_{\zeta}\right)=0$. At this stage, consider the set $\mathbb{D}^{\prime}_{\epsilon}$, which is given by the following form: $$\mathbb{D}^{\prime}_{\epsilon} =\left[0,1\right]\times\left[-2\pi,-\pi-\epsilon\right]\cup\left[-\pi+\epsilon,\pi-\epsilon\right]\cup\left[\pi+\epsilon,2\pi\right],$$
where $\epsilon \in \left.\right]0,\pi\left[\right.$. Notice that the following relations: $\mathfrak{U}\left(u,v,\pm 2\pi\right)=\mathfrak{U}\left(u,v,0\right)=0$, hold for every $u \in \left[0,1\right]$ and $v \in \left[0,1\right]$. The function $\mathfrak{U}\left(1,v,\Delta_{\zeta}\right)$ is continuous in the domain $\mathbb{D}^{\prime}_{\epsilon}$, as $\mathbb{D}^{\prime}_{\epsilon}\cap \mathbb{S}_1=\emptyset$ for every $\epsilon \in \left.\right]0,\pi\left[\right.$. The domain $\mathbb{D}^{\prime}_{\epsilon}$ is closed and bounded for every $\epsilon \in  \left.\right]0,\pi\left[\right.$. Therefore, the function $\mathfrak{U}\left(1,v,\Delta_{\zeta}\right)$ exhibits global maximum and global minimum in the domain $\mathbb{D}^{\prime}_{\epsilon}$, for every $\epsilon \in \left.\right]0,\pi\left[\right.$. These values occur at the stationary points which belong to the interior of the domain $\mathbb{D}^{\prime}_{\epsilon}$ or at the boundary of this domain.

As far as the boundary values of the function $\mathfrak{U}\left(u,v,\Delta_{\zeta}\right)$, with $v=1$, are concerned, we find that the following equation: $\partial_u\mathfrak{U}\left(u,1,\Delta_{\zeta}\right)=0$, exhibits the solution below,
\begin{eqnarray}
u=- Q\left(\frac{\omega_0}{T},\vartheta_0
\right)
\sec \Delta_{\zeta},\label{condVM2}
\end{eqnarray}
for every allowed value of the involved parameters. This solution fulfills the following equation: 
$\partial_{\Delta_{\zeta}}\mathfrak{U}\left(u,1,\Delta_{\zeta}\right)=0$. Again, the function $\mathfrak{U}\left(u,1,\Delta_{\zeta}\right)$ is continuous in the closed and bounded domain $\mathbb{D}^{\prime}_{\epsilon}$, for every $\epsilon \in \left.\right]0,\pi\left[\right.$. Hence, this function exhibits global maximum and global minimum in the domain $\mathbb{D}^{\prime}_{\epsilon}$, for every $\epsilon \in  \left.\right]0,\pi\left[\right.$. Again, these values occur at the stationary points which belong to the interior of the domain $\mathbb{D}^{\prime}_{\epsilon}$ or at the boundary of this domain.

The global maximum (\ref{VM}) and the global minimum (\ref{Vm}) of the short-time velocity are obtained by comparing the values which the functions $\mathfrak{U}\left(1,v,\Delta_{\zeta}\right)$ and $\mathfrak{U}\left(u,1,\Delta_{\zeta}\right)$ take in the corresponding stationary points and the global maximum and global minimum of the one-variable function $\mathfrak{U}\left(1,1,\Delta_{\zeta}\right)$. The evaluation of the latter values is straightforward and is reported below, for the sake of fluency. The procedure of comparison provides the global maximum (\ref{VM}). This value is obtained under the following conditions, which define the special NMSs $\mathfrak{S}_{\rm MV}$: 
\begin{eqnarray}
&&\hspace{-2.0em} \vartheta_1 =\frac{\pi}{2}, 
\hspace{1em}\vartheta_2 =\psi, \pi-\psi, \hspace{1em}\forall \,\, \Delta_{\zeta}\in \left[\right.
-\pi-\phi,-\pi\left[\right.\cup  \left[\right.
\pi-\phi,\pi\left[\right., \label{SNMSMaxIncr1}\\&&
\hspace{-2em}
\forall \,\, \vartheta_0\in 
\left[0,\frac{\pi}{2}\right.\Big[,\hspace{1em}\forall \,\, \zeta_0\in 
\big[0,2\pi\big[ \nonumber \\
&&\hspace{-2.0em} \vartheta_1 =\psi,\pi-\psi, \hspace{1em}\vartheta_2 =\frac{\pi}{2},  
\hspace{1em}
\forall \,\, \Delta_{\zeta}\in \left.\right]-\pi,
-\pi+\phi\left.\right]\cup  \left.\right]
\pi,\pi+\phi\left.\right]
\label{SNMSMaxIncr2},\\
&&
\hspace{-2em}\forall \,\, \vartheta_0\in 
\Big]\frac{\pi}{2},\pi\Big], \hspace{1em}\forall \,\, \zeta_0\in 
\big[0,2\pi\big[. \nonumber 
\end{eqnarray}
These conditions are obtained by considering the domain $\mathbb{D}^{\prime}_{\epsilon}$ for arbitrarily small, positive values of the parameter $\epsilon$. The following relations:
\begin{eqnarray}
&&\hspace{-2em} 0<- \frac{\sec \Delta_{\zeta}}{Q\left(\omega_0/T,\vartheta_0
\right)}\leq1, \label{cond1}\\
&&\hspace{-2em}0<- Q\left(\frac{\omega_0}{T},\vartheta_0
\right)
\sec \Delta_{\zeta} \leq 1,\label{cond2}
\end{eqnarray}
 are derived from Eqs. (\ref{condVM1}) and (\ref{condVM2}), by imposing the conditions $0<v\leq1$ and $0<u\leq1$, respectively. These relations define the angles $\phi$ and $\psi$, which are given by the forms below,
\begin{eqnarray}
&&\phi=\arccos \frac{1+w - \left|\cos \vartheta_0\right|\left(w-1\right)}{1+w + \left|\cos \vartheta_0\right|\left(w-1\right)}, \nonumber \\&&
\psi=\arcsin \frac{1+w - \left|\cos \vartheta_0\right|\left(w-1\right)}{\left|\cos \Delta_{\zeta}\right|\left[1+w + \left|\cos \vartheta_0\right|\left(w-1\right)\right] },\nonumber 
\end{eqnarray}
for every allowed value of the involved parameters. The above-mentioned procedure of comparison provides the global maximum (\ref{VM}). This value occurs under the following conditions which define the special NMSs $\mathfrak{S}_{\rm mV}$:
\begin{eqnarray}
&&\hspace{-1.0em} \vartheta_1 =\frac{\pi}{2}, \hspace{1em}\vartheta_2 =\psi,\pi-\psi, \hspace{1em}\forall \,\, \Delta_{\zeta}\in \left.\right]
-\pi,-\pi+\phi\left.\right]\cup  \left.\right]
\pi,\pi+\phi\left.\right], \label{SNMSMinIncr1}\\
&&\hspace{-1.0em}\forall \,\, \vartheta_0\in 
\left[0,\frac{\pi}{2}\right.\Big[ , \hspace{1.0em} 
\forall \,\, \zeta_0\in 
\big[0,2\pi\big[,
\nonumber \\ 
&&\hspace{-1.0em} \vartheta_1 =\psi,\pi-\psi, \hspace{1em}\vartheta_2 =\frac{\pi}{2}, \hspace{1em}\forall \,\, \Delta_{\zeta}\in \left[\right.-\pi-\phi,
-\pi\left[\right.\cup  \left[\right.
\pi-\phi,\pi\left[\right.,\label{SNMSMinIncr2}\\
&&\hspace{-1.0em}\forall \,\, \vartheta_0\in 
\Big]\frac{\pi}{2},\pi\Big], \hspace{1.0em} 
\forall \,\, \zeta_0\in 
\big[0,2\pi\big[. \nonumber  
\end{eqnarray}
Again, these conditions are determined by considering the domain $\mathbb{D}^{\prime}_{\epsilon}$ for arbitrarily small, positive values of the parameter $\epsilon$.

The boundary values of the function $\mathfrak{U}\left(u,v,\Delta_{\zeta}\right)$, with $u=v=1$, are analyzed by studying the form $\mathfrak{U}\left(1,1,\Delta_{\zeta}\right)$ as a function of the variable $r$ for every $r \in \left[-1,1\right]$, where $r \equiv \cos \Delta_{\zeta}$. The following equation: $\partial_r \mathfrak{U}\left(1,1,\Delta_{\zeta}\right)=0$, is solved by the forms below,
\begin{eqnarray}
&&r=\frac{- 1}
{Q\left(\omega_0/T,\vartheta_0
\right)}, \label{r1}\\
&&r=- Q\left(\frac{\omega_0}{T},\vartheta_0
\right).\label{r2} 
\end{eqnarray}
 The solution (\ref{r1}) is excluded for every $\vartheta_0 \in \left[ \pi/2, \pi \right]$, while the solution (\ref{r2}) is excluded for every $\vartheta_0 \in \left[ 0, \pi/2 \right]$. We remind that the value $\vartheta_0=\pi/2$ provides vanishing value of the function $\mathfrak{U}\left(u,v,\Delta_{\zeta}\right)$ in the domain $D\setminus \mathbb{S}$, and is, therefore, excluded. In this way, we find the special NMSs $\mathfrak{S}_{\rm MV}^{\prime}$, which provide the maximal velocity (\ref{VM}), and the special NMSs $\mathfrak{S}_{\rm mV}^{\prime}$, which induce the minimal velocity (\ref{Vm}). The special NMSs $\mathfrak{S}_{\rm MV}^{\prime}$, given by the conditions below, are particular cases of the special NMSs $\mathfrak{S}_{\rm MV}$,
\begin{eqnarray}
&&\hspace{-2.8em} \vartheta_1 =\vartheta_2=\frac{\pi}{2}, \hspace{1em}\Delta_{\zeta}=\pm \pi- \phi, \hspace{1em}\forall \,\, \vartheta_0\in \Big[0,\frac{\pi}{2}\Big[,\hspace{1em} \forall \,\, \zeta_0\in 
\big[0,2\pi\big[, \label{SNMSMaxIncr3}
\\&&\hspace{-2.8em} \vartheta_1 =\vartheta_2=\frac{\pi}{2}, \hspace{1em}\Delta_{\zeta}=\pm \pi+ \phi, \hspace{1em}\forall \,\, \vartheta_0\in \Big]\frac{\pi}{2},\pi\Big], \hspace{1em} \forall \,\, \zeta_0\in\big[0,2\pi\big[. \label{SNMSMaxIncr4}
\end{eqnarray}
Similarly, the special NMSs $\mathfrak{S}_{\rm mV}^{\prime}$, defined by the relations below, are particular cases of the special NMSs $\mathfrak{S}_{\rm mV}$, 
\begin{eqnarray}
&&\hspace{-2.8em} \vartheta_1 =\vartheta_2=\frac{\pi}{2}, \hspace{1em}\Delta_{\zeta}=\pm \pi+ \phi, \hspace{1em}\forall \,\, \vartheta_0\in \Big[0,\frac{\pi}{2}\Big[,  \hspace{1em} \forall \,\, \zeta_0\in\big[0,2\pi\big[,\label{SNMSMinIncr3}
\\&&\hspace{-2.8em} \vartheta_1 =\vartheta_2=\frac{\pi}{2}, \hspace{1em}\Delta_{\zeta}=\pm \pi- \phi, \hspace{1em}\forall \,\, \vartheta_0\in \Big]\frac{\pi}{2},\pi\Big],\hspace{1em} \forall \,\, \zeta_0\in\big[0,2\pi\big[.  \label{SNMSMinIncr4}
\end{eqnarray}

At this stage, we compare the values which the function $\mathfrak{U}\left(u,v,\Delta_{\zeta}\right)$ takes near any of the singular points, set $\mathbb{S}_1$, with the global maximum, $\mathfrak{V}_M\left(\vartheta_0, \omega_0/T\right)$, and the global 
minimum, $\mathfrak{V}_m\left(\vartheta_0, \omega_0/T\right)$. We find that the maximum and minimum values of the limits (\ref{VMexpr1}) and (\ref{VMexpr2}) reproduce the expressions (\ref{VM}) and (\ref{Vm}), respectively. The special NMSs (\ref{SNMSmaxQ1}), (\ref{SNMSminQ2}) and (\ref{SNMSmaxQ2}), (\ref{SNMSminQ1}) are obtained from the conditions under which the limits (\ref{VMexpr1}) and (\ref{VMexpr2}) are performed: $\Delta_{\zeta}\to \pm \pi$ for $u= Q\left(\omega_0/T,\vartheta_0\right)v$ and $\Delta_{\zeta}\neq\pm\pi$. This concludes the demonstration of the present results.


\begin{thebibliography}{0}

\bibitem{BP} H.-P. Breuer and F. Petruccione, \emph{The Theory of Open Quantum Systems}, Oxford University Press, Oxford (2002).

\bibitem{measure1}K. Kraus, \emph{States, Effects, and Operations}, Lecture Notes in Physics, Vol. {\bf 190}, Springer, Berlin (1983).

\bibitem{measure2}V.B. Braginsky and F. Ya Khalili, 
\emph{Quantum Measurements} (Cambridge University Press, Cambridge 1992).

\bibitem{measure3}P. Pechukas,  Phys. Rev. Lett. {\bf 73}, 1060 (1994).    

\bibitem{measure4}P. Stelmachovic and V. Buzek, Phys. Rev. A {\bf 64}, 062106 (2001).


\bibitem{measure5}H. Grabert, P. Schramm and G.L. Ingold,  Phys. Rep. {\bf 168}, 115 (1988).


\bibitem{measure6}L.D. Romero and J.P. Paz,  Phys. Rev. A {\bf 55}, 4070 (1997).

\bibitem{measure7} A.Z. Chaudhry and J. Gong, Phys. Rev. A {\bf 90}, 022101 (2014).

\bibitem{M0}V.V. Ignatyuk and V.G. Morozov, Phys. Rev. A {\bf 91}, 052102 (2015).
\bibitem{M1}V.V. Ignatyuk, Phys. Rev. A {\bf 92}, 062115 (2015).
\bibitem{M2} V.G. Morozov, S. Mathey and G. R\"opke, Phys. Rev A {\bf 85}, 022101 (2012).
\bibitem{M3} V.V. Ignatyuk and V.G. Morozov, Condens. Matter Phys. {\bf 16}, 34001 (2013).

\bibitem{G_IJQI2020}F. Giraldi, Int. J. Quantum Inform. {\bf 18}, 1941018 (2020).

\bibitem{LeggettRMP1987}A.J. Leggett \emph{et al.}, Rev. Mod. Phys. {\bf 59}, 1 (1987).


\bibitem{RH1} J. Luczka, Physica A {\bf 167}, 919 (1990).

\bibitem{GOSID2017} F. Giraldi, Open Syst. Inf. Dyn. { \bf 28}, 2150002 (2021), Open Syst. Inf. Dyn. { \bf 24}, 1750006 (2017); Open Syst. Inf. Dyn. { \bf 25}, 1850011 (2018).




\bibitem{SDPlenio1}M.P. Woods and M.B. Plenio, J. Math. Phys. {\bf 57}, 022105 (2016).

\bibitem{ReedSimonBook} M. Reed and B. Simon, {\it Methods of Modern Mathematical Physics Vol. 2}, Academics Press, Inc. (1975).


\end{thebibliography}
\end{document}